\documentclass[JHEP,12pt]{article}
\pdfoutput=1

\usepackage{graphics, color,soul}
\usepackage{graphicx}
\usepackage{amssymb}
\usepackage{jheppub}
\usepackage{lmodern,mathtools}

\usepackage{booktabs}
\usepackage[english]{babel}
\usepackage{amsmath,amssymb,amsbsy,amstext, amsthm, simplewick}
\usepackage{hyperref}
\usepackage{tikz}
\usepackage{xcolor}
\usepackage[normalem]{ulem}

\usepackage{amsfonts}
\usepackage{amssymb}
\usepackage{upgreek}
\usepackage{simplewick}
 \usepackage{exscale,relsize}
\usepackage{mathtools}
\usepackage{comment}

\usepackage[margin=1cm,labelfont={sf,bf,scriptsize},textfont={sf,scriptsize}]{caption}

\def\cJ{\mathcal{J}}

\def\cO{\mathcal{O}}

\def\cJ{\mathcal{J}}

\def \la {\langle}
\def \ra {\rangle}
\def \p {\partial}

\def \b {\bar}

\def\nn{{\nonumber}}

\def\be#1\ee{\begin{align}#1\end{align}}

\def\citeHLLH{Kulaxizi:2018dxo,Fitzpatrick:2019zqz,Karlsson:2019qfi,Kulaxizi:2019tkd,Fitzpatrick:2019efk,Karlsson:2019dbd,Li:2019zba,Li:2020dqm,Parnachev:2020fna,Fitzpatrick:2020yjb,Parnachev:2020zbr,Karlsson:2021duj,Karlsson:2021mgg}

\definecolor{blueM}{RGB}{0,87,184}
\definecolor{yellowM}{RGB}{254,221,0}

\begin{document}

\begin{flushleft}
 \hfill \parbox[c]{40mm}{CERN-TH-2022-065}
\end{flushleft}

\title{\href{https://theory.cern/ukraine}{\color{blueM} Gravitational orbits, double-twist mirage, and \color{yellowM} many-body scars}}

\author{Matthew Dodelson and}
\author{Alexander Zhiboedov}

\abstract{We explore the implications of stable gravitational orbits around an AdS black hole for the boundary conformal field theory. The orbits are long-lived states that eventually decay due to gravitational radiation and tunneling. They appear as narrow resonances in the heavy-light OPE when the spectrum becomes effectively continuous due to the presence of the black hole horizon. Alternatively, they can be identified with quasi-normal modes with small imaginary part in the thermal two-point function. The two pictures are related via the eigenstate thermalisation hypothesis. When the decay effects can be neglected the orbits appear as a discrete family of double-twist operators. We investigate the connection between orbits, quasi-normal modes, and double-twist operators in detail. Using the corrected Bohr-Sommerfeld formula for quasi-normal modes, we compute the anomalous dimension of double-twist operators. We compare our results to the prediction of the light-cone bootstrap, finding perfect agreement where the results overlap. We also compute the orbit decay time due to scalar radiation and compare it to the tunneling rate. Perturbatively in spin, in the light-cone bootstrap framework double-twist operators appear as a small fraction of the Hilbert space which violate the eigenstate thermalization hypothesis, a phenomenon known as many-body scars.  Nonperturbatively in spin, the double-twist operators become long-lived states that eventually thermalize. We briefly discuss the connection between perturbative scars in holographic theories and known examples of scars in the condensed matter literature.
}

\affiliation{CERN, Theoretical Physics Department, CH-1211 Geneva 23, Switzerland}

 \maketitle
\section{Introduction}
Gravitating bodies in asymptotically flat spacetimes admit stable particle orbits in four dimensions. These provide some of the simplest experimental tests of general relativity (and basis for our everyday experience). For example, the precession of the perihelion of Mercury is famously explained by geodesic motion in the Schwarzschild geometry. When gravitational radiation is taken into account, one discovers that these orbits are in fact metastable, since they can lose energy to the gravitational field and eventually fall into the black hole. The gravity waves emitted from this inspiral process were observed by the LIGO/Virgo collaboration, leading to another powerful confirmation of general relativity \cite{LIGOScientific:2016aoc}.  \\
\indent Given that orbits and gravity waves play such a fundamental role in asymptotically flat spacetime, it is interesting to investigate their significance in the asymptotically AdS case as well \cite{Festuccia:2008zx,Berenstein:2020vlp}. From the AdS/CFT point of view, gravitational orbits are interesting for the following simple reason. On one hand, quantum-mechanical systems with simple gravity duals are known to be maximally chaotic \cite{Maldacena:2015waa}. On the other hand, gravitational orbits present an example of dynamics which is easily reversible in time and in this sense is integrable. This paradoxical aspect of gravitational theories is similar to the fact that when chaos becomes maximal the scattering becomes purely elastic \cite{Shenker:2014cwa,Gu:2018jsv,Costa:2012cb}. Also, due to the existence of gravity waves, we might expect the orbits to eventually fall into the black hole, just like their flat space counterparts. In the quest of identifying the quantum-mechanical systems with simple gravity duals it is therefore very important to identify the salient features related to both chaos and approximate integrability of the dual gravitational dynamics. The same comment applies to flat space and de Sitter holography as well. 

In this paper we explore some basic aspects of the AdS orbital dynamics, both from the bulk and boundary perspectives, extending the analysis of \cite{Festuccia:2008zx,Berenstein:2020vlp}. At the classical level, gravitational orbits are associated to a set of Regge trajectories in the dual CFT.\footnote{We use the term ``Regge trajectory'' to specify a family of states with scaling dimension $\Delta_n(J)$ labeled by spin $J$ and potentially some other quantum numbers $n$ whose dependence is analytic in $J$.} These are characterized by two quantum numbers: spin $J$ (related to the size of an orbit) and radial excitation $n$ (related to the eccentricity of an orbit). There are two basic features that distinguish $AdS_{d+1}$ orbits from their flat space counterparts: stable orbits exist in any $d \geq 3$,\footnote{Stable orbits exist below the BTZ threshold in $AdS_3$ as well \cite{Fitzpatrick:2014vua}.} and the energy and angular momentum of an orbit around a black hole grow with the orbital radius. There are two basic mechanisms that render AdS orbits unstable: emission of gravity waves and tunneling of the orbiting body into the black hole. Emission of gravity waves is very universal and it does not rely on the precise nature of the gravitating bodies (whether they are black holes or stars). In this paper we will not directly analyze gravity waves, but we will instead consider the closely related problem of computing the lifetime of an orbit due to the emission of scalar radiation in the semi-classical approximation. An important aspect of gravitational radiation is that it is ${1 \over c_T}$ suppressed and as such is not present in the large $c_T$ limit.\footnote{Here $c_T$ stands for the two-point function of the stress-energy tensor. Since we assume a simple gravity dual we also assume that the 't Hooft coupling $\lambda$, or gap in the spectrum of higher spin single trace operators \cite{Heemskerk:2009pn}, is large in this paper. } Tunneling, on the other hand, is present in the large $c_T$ limit as well but it requires the presence of the black hole horizon.

\indent What is the nature of the orbit states on the AdS boundary? The simplest example of this type corresponds to a binary system of light operators supported by the centrifugal potential in AdS. The boundary dual of this system is provided by a family of double-twist operators $[\mathcal{O}_L\mathcal{O}_L]_{n,J}$, where $n$ and $J$ are the same quantum numbers that appeared above, and $\mathcal{O}_L$ denotes a light operator \cite{Alday:2007mf,Fitzpatrick:2012yx,Komargodski:2012ek}. Such states are completely universal and persist as energy eigenstates at finite $c_T$ as well, even though their precise bulk nature at finite $c_T$ is not known.\footnote{Presumably, they correspond to orbiting states dressed by a specific cloud of gravitational radiation that make the whole system an energy eigenstate.} 

Consider next a situation where one of the operators is heavy, $\Delta_H \sim c_T$, such that its AdS dual is a black hole. By a naive analogy we can try to associate a set of double-twist operators $[\mathcal{O}_H\mathcal{O}_L]_{n,J}$ to orbits in this case as well. And, indeed, such operators were recently observed in the light-cone bootstrap analysis of the heavy-light four point function \cite{\citeHLLH}. Their presence in the spectrum, however, is puzzling from the bulk perspective. Indeed, due to tunneling we expect orbits to be meta-stable already to leading order in $c_T$. Relatedly, due to the presence of the black hole horizon in the bulk we expect the CFT spectrum to be effectively continuous and not be given by a discrete set of double-twist operators. The correct picture in this case is that orbit states are narrow resonances: they present pole singularities of the conformal partial waves $c(\Delta,J)$, see e.g. \cite{Mack:2009mi,Costa:2012cb,Caron-Huot:2017vep}, on the second sheet.\footnote{At finite $c_T$, $c(\Delta,J)$ are meromorphic functions. However, they develop a cut in the large $c_T$ limit in the presence of a heavy state dual to a black hole. We can then study the multi-sheeted structure of $c(\Delta,J)$.} More physically, they represent a superposition over a small band of energy eigenstates.

A more familiar manifestation of the same phenomenon is known as quasi-normal modes \cite{cardoso}. Quasi-normal modes are defined as normalizable solutions to bulk wave equations which are purely ingoing at the horizon. They lead to poles in the retarded thermal two-point function. At non-zero spin they are directly related to orbits as shown in \cite{Festuccia:2008zx}. As we will review, via the eigenstate thermalization hypothesis (ETH) \cite{srednicki1999approach,DAlessio:2015qtq,Lashkari:2016vgj,Delacretaz:2020nit} they become poles on the second sheet of $c(\Delta,J)$.

\indent This picture of orbit states as resonances raises an immediate question for the light-cone bootstrap program in the heavy-light regime, which aims to compute the operator product coefficients and anomalous dimensions of double-twist operators $[\cO_H \cO_L]_{n,J}$. If these operators are not energy eigenstates, then what is actually being computed by the bootstrap? The resolution of the puzzle comes from the fact that there is the following equivalence
\be
\label{eq:mirage}
 {1 \over 2 \pi i} \Big( {1 \over \Delta - \Delta_n(J) - i e^{- c_0(\mu) J}} - {1 \over \Delta - \Delta_n(J) + i e^{- c_0(\mu) J}}\Big) \overset{\text{PT}}{\simeq} \delta(\Delta - \Delta_n(J) )   ,
\ee
where $\mu \sim {\Delta_H \over c_T}$ and $\text{PT}$ stands for perturbatively in $1/J$.\footnote{Let us emphasize that we are talking about the continuum emerging in the large $c_T$ limit. In particular, $e^{- c_0(\mu)J} \gg e^{- c_T}$, where $e^{- c_T}$ is the scale associated with the discreteness of the CFT spectrum.} Here the RHS corresponds to the spectral density with an isolated operator of dimension $\Delta_n(J)$. The LHS represents, on the other hand, a continuous spectrum containing a pair of narrow resonances whose width is nonperturbative in spin $J$ and is thus not visible in perturbation theory. This explains why \cite{\citeHLLH} have observed a discrete spectrum, even though the true spectrum is continuous.

Using the ETH we derive the expansion of the correlator in terms of the quasi-normal modes (QNMs). We then notice that perturbatively in ${1 \over J}$ and leading order in $c_T \to \infty$ it becomes the ordinary OPE expansion, where the sum over QNMs becomes the sum over the double-twist operators.
In this sense the discrete spectrum of heavy-light double-twist operators is a {\it mirage} that emerges in the large spin perturbation theory. In the light-light channel large $J$ perturbation theory is captured by the multi-trace stress-energy tensor operators, schematically $T^n$.
Note that this simple analysis shows that the conclusion of \cite{Fitzpatrick:2012yx,Komargodski:2012ek} regarding the existence of a discrete family of double-twist operators is not a consequence of crossing symmetry in the presence of the continuum spectrum, or, equivalently, for $c_T$ being the largest parameter in the problem.

\indent The identification of double-twist operators with bulk orbit states is not just a conceptual point. We show that it provides a powerful tool for computing anomalous dimensions to all orders in $\mu$ at large $\Delta_L$ and large spin $J$. Indeed, as explained above, the metastable orbits are in one-to-one correspondence with long-lived quasi-normal modes with large spin. These modes are subject to the Bohr-Sommerfeld quantization rule, which is applicable when the mass of the orbiting particle, or equivalently $\Delta_L$, is large \cite{Festuccia:2008zx}. Moreover, we will see that corrections in $1/\Delta_L$ can be systematically computed by analyzing corrections to Bohr-Sommerfeld. This allows us to match known results from the light-cone bootstrap literature, and extend these results to all orders in $\mu$.\\
\indent After computing the spectrum of resonances, we analyze correlation functions in the orbit states. In particular, it is interesting to ask whether these states behave like typical high-energy states, or if there are simple measurements that can be done to distinguish them from black hole microstates. Using the light-cone bootstrap at five points, as in the recent work \cite{Antunes:2021kmm}, we show that light operators have one-point functions of order one in the orbit states. In contrast, a one-point function in a typical black hole microstate is suppressed as $\lambda,c_T\to\infty$. If the orbit states were exact energy eigenstates, this would imply a violation of ETH, since the one-point function would not be a smooth function of energy. \\
\indent This apparent violation of ETH is not necessarily a contradiction. ETH-violating states are known as many-body quantum scars, and many examples can be found in the condensed matter literature starting from \cite{bernien2017probing,turner}, see e.g. \cite{serbyn,moudgalya2021quantum} for reviews. In our case, the orbit states are not true scars, since they eventually decay. However, we may think of them as scars perturbatively in ${1 \over J}$ and at infinite $c_T$, where the decay rate is zero. We also comment on the role of bulk higher spin symmetry in organizing the spectrum of orbit states at large $J$.

The plan of the paper is the following. In Section \ref{sec:classicalorbits}, we review properties of stable orbits around AdS-Schwarzschild black holes. In Section \ref{sec:raddecay} we analyze the decay of the orbits due to the emission of radiation. In Section \ref{sec:orbitsLCbootstrap} we consider the heavy-light four-point function and use the ETH to write the heavy-light OPE expansion in terms of the QNMs. We then discuss how it reduces to the sum over double-twist operators in perturbation theory. In Section \ref{sec:doubletwistdim} we use the Bohr-Sommerfeld formula and corrections to it to compute the anomalous dimensions of the double-twist operators.  In Section \ref{sec:scars} we explore the connection between the gravitational orbits and many-body scars. We end with conclusions and a few open directions.

\section{Classical orbits}
\label{sec:classicalorbits}

In this section we analyze classical stable orbits around AdS black holes. We focus on the case of $AdS_4$, which is the minimal number of spacetime dimensions that admits stable orbits around black holes. It has an additional virtue of admitting stable orbits in the flat space regime as well. We show that classical gravitational orbits are naturally associated to the double-twist-like Regge trajectories in the boundary CFT \cite{Berenstein:2020vlp} --- we will explore this connection in great detail in the next sections.

Higher-dimensional $AdS_{d+1}$ cases are completely analogous and we also briefly discuss them in this section. Finally, we comment on orbits in the presence of higher-derivative corrections and orbits in dS. 

\subsection{Review of stable orbits}
\indent We consider the Schwarzschild-AdS black hole in four dimensions. We focus on classical, stable time-like orbits in this geometry \cite{Berenstein:2020vlp}, which we review next. 

The black hole metric in Schwarzschild coordinates takes the form \cite{Witten:1998zw}
\begin{align}
ds^2=-f(r)\, dt^2+\frac{dr^2}{f(r)}+r^2\, d\Omega^2, \hspace{10 mm}f(r)={r^2 \over R_{AdS}^2}+1-\frac{GM}{r},\label{schmetric}
\end{align}
where $R_{AdS}$ is the AdS radius and $M$ is proportional to the mass of the black hole. The black hole horizon is located at $f(r_s)=0$ and the AdS boundary is at $r=\infty$.

Next we consider a probe, classical body that follows a timelike geodesic in the black hole geometry. Due to the symmetries of the problem, the geodesic motion is characterized by the conserved energy $E$ and the angular momentum $L$ per unit mass. In terms of these quantities the equation of motions take the form
\begin{align}
\dot{t}&=\frac{E}{f(r)}, ~~~\dot{\phi}=\frac{L}{r^2},~~~\dot{r}^2=E^2-V(r),
\label{eq:timelike}
\end{align}
where we introduced the potential 
\be
\label{eq:potentialBH}
V(r)=f(r)\left(\frac{L^2}{r^2}+1\right) .
\ee
From \eqref{eq:potentialBH} it is clear that the potential is zero at the horizon $V(r_s)=0$, and goes to infinity at the AdS boundary.

It is convenient to introduce a dimensionless parameter $\mu$
\be
\mu \equiv {G M \over R_{AdS}} ,
\ee
and measure distances and other dimensionful quantities in AdS units by setting 
\be
R_{AdS}=1 .
\ee
With that in mind, the flat space limit corresponds to $\mu \ll 1$, whereas large 
black holes that dominate the canonical ensemble in the dual CFT correspond to $\mu>2$ \cite{Hawking:1982dh,Witten:1998zw}.
\vspace{0.1cm}
\subsubsection{  Circular orbits}
\vspace{0.1cm}

First, we consider the simplest case of circular orbits at constant radial distance $r_0$. They are found by finding a critical point of the potential $V'(r_0)=0$. This condition can be solved as follows,
\begin{align}
\label{eq:orbitsQN}
E&=\frac{f(r_0)}{\sqrt{1-3 \mu/(2r_0)}}, ~~~~~ L=r_0^2 \frac{\sqrt{1+\mu/(2 r_0^3)}}{\sqrt{1-3 \mu/(2r_0)}} \ .
\end{align}
The circular orbits only exist outside the photon sphere\footnote{The photon sphere is the location of unstable, circular, null geodesics in the black hole geometry.} $r>3\mu/2$, as can be seen from \eqref{eq:orbitsQN} by requiring positivity of the square root in the denominator. 

We will be only interested in the stable orbits in this paper\footnote{Although the unstable orbits are not directly relevant for our analysis, they lead to interesting singularities in the two-point function at finite temperature \cite{Hubeny:2006yu,Dodelson:2020lal}.}. These correspond to the minimum of the potential or, equivalently, $V''(r_0)>0$. The stability condition takes the form
\be
\left( r - 3 \mu \right) + 8 r^3  \left({r \over \mu} - {15 \over 8} \right) > 0 .
\ee
Let us note that the orbits with $r > 3 \mu $ are stable for any value of the dimensionless parameter $\mu$. In the flat space limit $\mu \to 0$, $r = 3 \mu$ becomes an inner-most stable orbit (ISCO). In the opposite limit of large black holes in AdS, $\mu \to \infty$, we get that stable orbits exist for $r > {15 \over 8} \mu$. Orbits with $3 \mu>r>{15 \over 8} \mu$ are stable depending on the precise value of $\mu$.

Let us now discuss the description of this state in terms of the CFT dual. A particle of mass $m_L$ in AdS corresponds to an operator with scaling dimension $\Delta_L \simeq m_L R_{AdS}$, where we assumed that $\Delta_L \gg 1$. This assumption effectively makes the bulk particle classical and the analysis of the present section accurate. The angular momentum $J$ in the CFT is related to $L$ as follows
\be
J =\Delta_{L} L . 
\ee 
Similarly, $E$ measures energy per unit mass. In this way circular orbits describe the following Regge trajectory in the dual CFT,
\be
\Delta_{H,L}(J) &= \Delta_H + \Delta_L E \nn \\
&= \Delta_H + \Delta_L + J + \gamma(\mu , J) , ~~~ J \geq J_{\text{min}}(\mu) ,
\ee
where $\Delta_H$ is related to the mass of the black hole, $\Delta_L$ is related to the mass of the probe,  and $J_{\text{min}}(\mu)$ is the minimal spin for which the stable orbit exists. For $\mu \ll 1$, $J_{\text{min}}(\mu) \sim \sqrt{3}\Delta_L \mu$, whereas for $\mu \gg 1$ we have $J_{\text{min}}(\mu) \sim {225 \over 64} \sqrt{5}\Delta_L \mu^2$.

The anomalous dimension takes the form
\be
\gamma(\mu, J) = \Delta_L \Big(E- L -1 \Big) \Big|_{L = {J \over \Delta_L}} .
\ee
This formula is correct for fixed $\mu$ and $L= {J \over \Delta_L}$ and to leading order in the $\Delta_L \to \infty$ limit. In other words, the formula above has $\Delta_L^{-1}$ corrections due to the quantum fluctuations of the particle around the classical orbit which we will discuss further below. 

\begin{figure}[t]
  \hspace{-24pt}   
  \begin{tabular}{c c}
\includegraphics[scale=0.7]{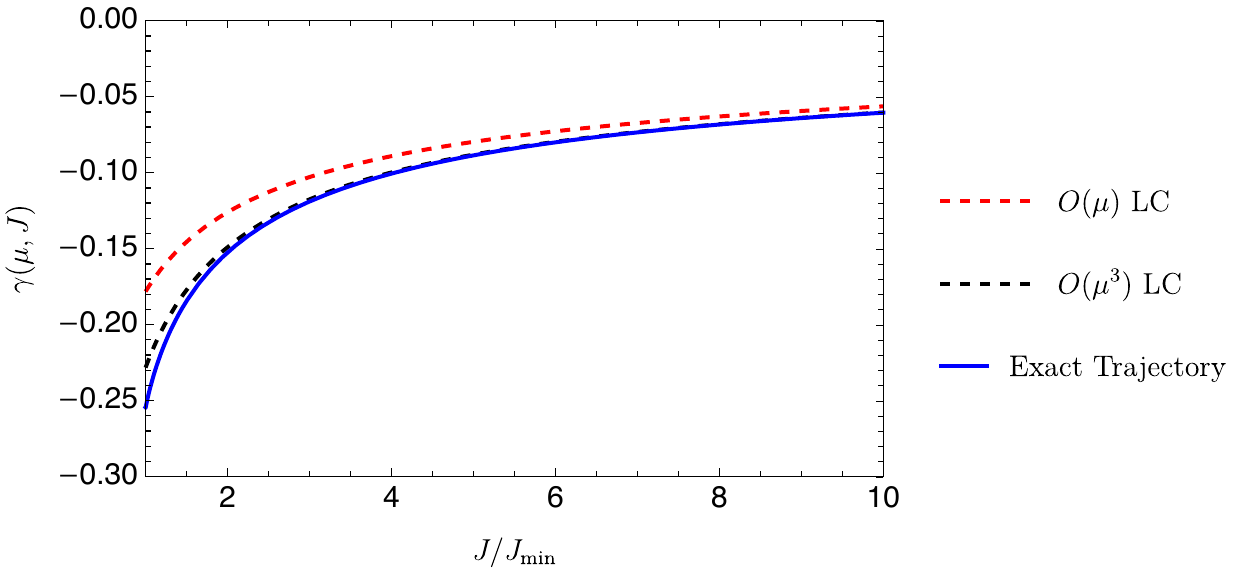} & \includegraphics[scale=0.5]{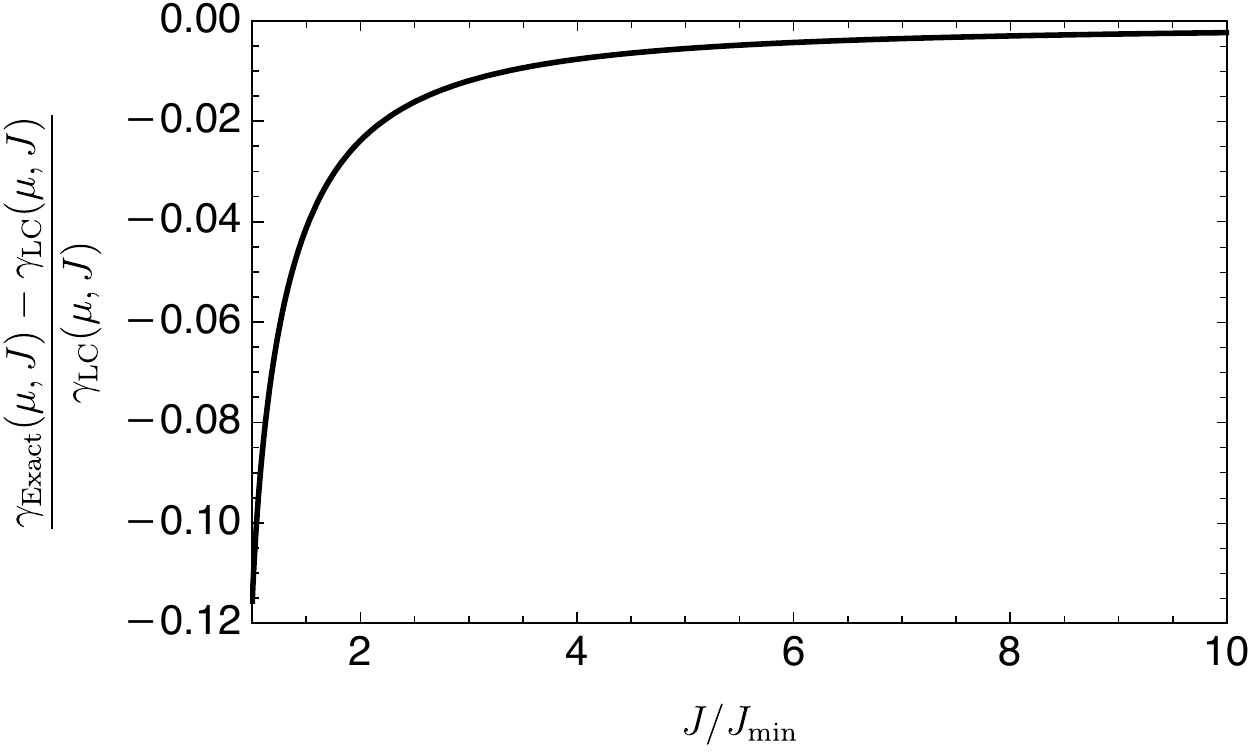}\\
(a) & (b)                         
 \end{tabular}
 \caption{The exact Regge trajectory for circular orbits versus its large spin expansion approximation. We set $\mu=5$ for which $J_{\rm min} \simeq 196.83 \Delta_L $. For other values of $\mu$ the situation is very similar. (a) We plot the exact Regge trajectory against its large spin approximation given by \eqref{eq:largespin4d}. The red dashed curve corresponds to only keeping the leading $O(\mu)$ term in \eqref{eq:largespin4d}. The black dashed curve corresponds to keeping all terms in \eqref{eq:largespin4d}. The exact Regge trajectory in blue can be plotted using its parameteric representation given by \eqref{eq:orbitsQN}. (b) The relative error in approximating the exact anomalous dimension $\gamma_{\text{Exact}}$ by its large spin expansion up to order $O(\mu^3)$ as in \eqref{eq:largespin4d}, which we denote by  $\gamma_{\text{LC}}$. We see that the error is of order $10 \%$ at $J= J_{\text{min}}$, and becomes less than $1 \%$ for $J \gtrsim 3 J_{\text{min}}$.  }
  \label{fig:ReggeTr}
\end{figure}

For convenience let us write down the first few terms in the expansion at small $\mu$,
\be
\label{eq:largespin4d}
{\gamma(\mu, J) \over \Delta_L} =-\frac{\mu }{2 \sqrt{L}} - \frac{9}{32} {\mu ^2 \over L} \left( 1 + \frac{1}{9 L}\right) -  \frac{81 \mu^3}{256 L^{3/2}} \left(1+ {1 \over 9 L}\right)^2 + ... \Big|_{L = {J \over \Delta_L}} \ . 
\ee
Several comments are in order. First, we note that the small $\mu$ expansion is closely related to the large $L$ expansion. Moreover, the large $L$ expansion translates into the large spin $J$ expansion which naturally appears in the context of the light-cone bootstrap \cite{\citeHLLH}. Finally, we observe that all terms in the expansion are sign-definite. The same pattern continues when higher orders in $\mu$ are included. It is interesting to plot the exact Regge trajectory against its large spin expansion. We present the result in Figure \ref{fig:ReggeTr}.

Let us next discuss the flat space limit of the orbits above. In dimensionless units this limit corresponds to taking $\mu \to 0$, while keeping $\Delta_L \mu \simeq G M m_L $ and $J = L \Delta_L$ fixed. In particular, this implies that $L \sim \mu$ in the flat space limit. It is convenient to take the limit at the level of \eqref{eq:orbitsQN}. In this way we get the following Regge trajectory,
\be
\Delta_{H,L}^{\text{flat}}(J) &= \Delta_H + \Delta_L + \gamma_{\text{flat}}(\Delta_L \mu , J) , ~~~ J \geq \sqrt{3} \Delta_L \mu ,
\ee
where the Regge trajectory takes the form
\be
{\gamma_{\text{flat}}(\Delta_L \mu , J) \over \Delta_L} &= \frac{\sqrt{2}}{3}  \sqrt{\frac{1}{ \sqrt{1-\frac{3 (\Delta_L \mu)^2}{J^2}}+1}+\sqrt{1-\frac{3 (\Delta_L \mu)^2}{J^2}}+3}-1 \nn \\
&\simeq -\frac{\mu ^2 \Delta _L^2}{8 J^2} -\frac{9 \mu ^4 \Delta _L^4}{128 J^4} - ... \ .
\ee
Note the absence of the term $+J$ in the formula for the Regge trajectory, which would be present in AdS. This is a manifestation of the familiar fact that the binding energy of a flat space orbit decays as a function of spin.

\vspace{0.1cm}
\subsubsection{ Non-circular orbits}
\vspace{0.1cm}

\begin{figure}[t]
\includegraphics[scale=.7]{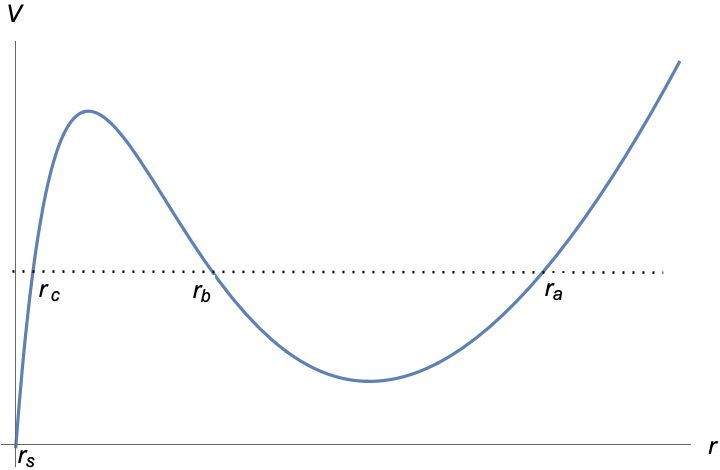}
\centering
\caption{For $J>J_{\text{min}}(\mu)$, the potential $V(r)$ goes to zero at the horizon $r=r_s$, behaves like $r^2$ for large $r$, and has a metastable minimum at the radial position of the circular orbit. For a given value of $E$, there are three turning points $r_a$, $r_b$, and $r_c$ where $E^2=V(r)$, with $r_a>r_b>r_c$.\label{potentialfig}}
\centering
\end{figure}

\indent We can also consider non-circular orbits, see Figure \ref{potentialfig}. In this case $r$ does not stay constant but changes between $r_b \leq r \leq r_a$. It is therefore natural to characterize non-circular orbits by eccentricity $x$, defined 
as follows
\be
\sqrt{1-x^2} \equiv r_b/r_a . 
\ee From the condition $V(r_a) = V(r_b)$ we get for the energy and angular momentum
\be
\label{eq:noncircQN}
E=\sqrt{V(r_a)}, ~~~~~ L=r_a^2 \frac{y \sqrt{1+\mu/(y (1+y) r_a^3)}}{\sqrt{1-\mu(1+y+y^2)/(y(1+y)r_a)}} ,
\ee
where we introduced $y = \sqrt{1-x^2}$. By setting $y=1$, or $x=0$, we reproduce the previous formula \eqref{eq:orbitsQN}.

A convenient way to think about the non-circular orbits is the following. Let us fix $\mu$ and $L$, which fixes the form of the potential. We then consider energy levels of a particle in this potential labeled by $n$.
The quantization condition takes the form
\be
\label{eq:BSclassical}
\Delta_L \int_{r_b}^{r_a} {d r \over f(r)} \sqrt{E^2 - V(r)} = \pi n ,
\ee 
where we are only interested in the terms that contribute to leading order in the classical $\Delta_L \gg 1$ limit. In particular, classical non-circular orbits correspond to $n \gg 1$ such that ${n \over \Delta_L}$ is kept fixed. We will discuss various corrections to \eqref{eq:BSclassical} in Section \ref{sec:doubletwistdim}.

Given $L$ and $r_a$, \eqref{eq:BSclassical} fixes $n$. And conversely, given $L$ and $n$ the corresponding orbit labeled by $y$ and $r_a$ can be identified via \eqref{eq:noncircQN} and \eqref{eq:BSclassical}. Note that if we fix $L$ and increase $n$ the corresponding orbits correspond to larger and larger values of $E$. By setting in \eqref{eq:BSclassical} $E^2 = V_{\text{max}}(\mu,L)$ we find that $n$ is bounded above,
\begin{align}
n\le n_{\text{max}}(\mu,J).
\end{align}
It is not hard to check that for $\mu \ll 1$ and $L\gg \mu$ we have $V_{\text{max}}(\mu,L) \simeq \frac{4L^2}{27\mu^2}$. Correspondingly, we have ${n_{\text{max}}(\mu,J) \over \Delta_L} \sim {4 L^2 \over 27 \pi \mu^2}$.

As before, we can associate to non-circular orbits a family of Regge trajectories in the dual CFT. In terms of the quantum numbers introduced above, the dimension of the dual state takes the form
\be
\Delta_{H,L}(n, J) = \Delta_H + \Delta_L + 2n + J + \gamma(\mu , n, J) ,~~~ J \geq J_{\text{min}}(\mu), ~~~ n\leq  n_{\text{max}}(\mu,J), 
\ee
where as before
\be
\label{eq:anomdimn}
\gamma(\mu, n, J) &= \Delta_L \Big(E- L -1 - {2n \over \Delta_L} \Big)  \Big|_{L = {J \over \Delta_L}} \nn \\
&=\sum_{k=1}^{\infty} \mu^k \gamma_k (n, J) ,
\ee
and as before $\gamma_k (n, J) \to 0$ at large $J$. We will provide more details on the explicit form of $\gamma_k (n, J)$ below. 

From the formulas above it is not obvious that the term $\mu^0$ is absent in \eqref{eq:anomdimn}. It is instructive to demonstrate this explicitly and we do so in Section \ref{sec:doubletwistdim}.

\subsection{Multi-orbit states}

In the sections above we discussed the simplest case of an orbit where we consider a single orbiting body around the black hole. In the same way we could have considered multi-orbit states and analyzed their properties. These should be related to the multi-twist operators in the dual theory as in \cite{Fitzpatrick:2012yx,Komargodski:2012ek}.

In particular, we can imagine the Milky Way galaxy in the middle of AdS with the black hole Sagittarius A* of mass $\mu_{A^*}$ at its center. Its dynamics will be encoded in the complicated properties of the multi-twist QNMs in the CFT dual. These would look like multi-twist operators perturbatively at large $J$.  

\subsection{General spacetime dimensions}
Let us briefly discuss the situation in general number of spacetime dimensions $d$. In this case the red shift factor in the metric for the $AdS_{d+1}$ black hole takes the form \cite{Witten:1998zw}
\be
f_d(r)={r^2 \over R_{AdS}^2}+1-\frac{GM}{r^{d-2}}.
\ee

\noindent First, for $d=2$, in $AdS_3$ the criticality of the potential $V'(r_0)$ takes the following form
\be
r_0^4 = R_{AdS}^2 L^2 (1 - G M) . 
\ee
Therefore orbits exist only below the BTZ threshold $GM<1$. One can also easily check that these orbits are stable \cite{Fitzpatrick:2014vua}. 
These orbits disappear in the flat space limit $R_{AdS} \to \infty$ limit.

For $d>2$, timelike circular orbits exist around the Schwarzschild black hole in $AdS_{d+1}$ only for
\be
r_0 > \Big( {d G M \over 2} \Big)^{{1 \over d-2}} . 
\ee
Their energy and angular momentum take the form
\begin{align}
\label{eq:orbitsQNd}
E&=\frac{f_d(r_0)}{\sqrt{1-dGM/(2r_0^{d-2})}}, ~~~~~ L={r_0^2 \over R_{AdS}} \frac{\sqrt{1+(d-2)GM R_{AdS}^2/(2 r_0^d)}}{\sqrt{1-d GM/(2r_0^{d-2})}}\ .
\end{align}

The stability of orbits condition $V''(r_0)>0$ takes the form
\be
(d-2) GM R_{AdS}^2 \Big( (4-d) r_0^{d-2} - d G M \Big) + 8 r_0^{d} \Big( r_0^{d-2} - {d (d+2) \over 8} G M \Big) > 0 .
\ee
Let us consider this condition in the flat space limit $R_{AdS} \to \infty$. We see that no stable orbits exist for $d \neq 3$. This is the famous fact about celestial motion in flat space. In the opposite limit $r_0^{d-2} \sim GM \gg (R_{AdS})^{d-2}$, on the other hand, stable orbits always exist and the story is very similar to the case of $AdS_4$ considered at the beginning of this section. 

\subsection{Higher derivative corrections}
We can also consider higher derivative corrections to the Regge trajectories, which are small at large 't Hooft coupling $\lambda$ \cite{Camanho:2014apa,Afkhami-Jeddi:2016ntf,Afkhami-Jeddi:2017rmx,Kulaxizi:2017ixa,Costa:2017twz,Caron-Huot:2022ugt}. As a particular example, we take the case of Einstein gravity with a Gauss-Bonnet term,
\begin{align}
S=\frac{1}{16\pi G}\int d^{d+1}x\, \sqrt{-g}\left(R+\frac{d(d-1)}{l^2}+\alpha(R_{\mu\nu\lambda\delta}R^{\mu\nu\lambda \delta}-4R_{\mu\nu}R^{\mu\nu}+R^2)\right).
\end{align}
In addition to purely gravitational higher derivative terms, there can be couplings between matter and curvature such as $\phi W^2$, where $W$ is the Weyl tensor. However, for $\Delta_L\gg 1$ such terms are subleading, since the mass term $\sqrt{-g}\Delta_L^2\phi^2$ in the potential for $\phi$ dominates.\footnote{However, the coupling $\phi W^2$ induces a nontrivial one-point function \cite{Grinberg:2020fdj}.}\\
\indent In $d=3$ the Gauss-Bonnet term is topological, but in higher dimensions it is nontrivial and a spherically symmetric black hole solution exists \cite{Boulware:1985wk,Cai:2001dz}. The redshift factor is
\begin{align}
f(r)=1+\frac{r^2}{2\alpha}\left(1-\sqrt{1+4\alpha\left(\frac{ GM}{r^{d}}-\frac{1}{l^2}\right)}\right).
\end{align}
This is an asymptotically AdS spacetime, whose AdS radius is related to $l$ by 
\begin{align}
l=\frac{R^2_{AdS}}{\sqrt{R_{AdS}^2-\alpha}}
\end{align}
\indent We now take $\alpha$ to zero, holding $R_{AdS}$ fixed. We
set $R_{AdS}=1$ and $\mu=GM$. Solving for the stable orbits gives
\begin{align}
E&=E(\alpha=0)+\alpha\frac{\mu r_0^{2-d/2}(r_0^d(dr_0^2+d-4)-(d-2)\mu)}{2(r_0^d-d\mu r_0^2/2)^{3/2}}+O(\alpha^2) \\
L&=L(\alpha=0)+\alpha \frac{\mu r_0^2(2r_0^d(dr_0^2+d-2)-2\mu(dr_0^2+d-1)+d\mu^2 r_0^{2-d})}{(2r_0^d-d\mu r_0^2)^{3/2}\sqrt{2r_0^d+(d-2)\mu }}+O(\alpha^2).
\end{align}
Solving for the anomalous dimension perturbatively in $\mu$, we find

\begin{align}
\label{eq:gendgammaorbit}
\frac{\gamma(\mu,J)}{\Delta_L}&=-\frac{1+2\alpha}{2} \frac{\mu}{L^{d/2-1}}-{d^2 \over 32} \left(\frac{\mu}{L^{d/2-1}}\right)^2  \Big(1+4 \alpha + {(1+4 \alpha) (d-4) + {4 \over d} \over d L} \Big) \\
&\hspace{5 mm}-\frac{d^4}{256}\left(\frac{\mu}{L^{d/2-1}}\right)^3\Big( \left(1+{(d-2)^2 \over d^2 L} \right)^2(1+6\alpha)-\frac{32 \alpha}{d^2 L} \left( 1+\frac{(d-2) (d-1)}{d^2 L} \right) \Big) \nn \\ 
&\hspace{5 mm}+O(\mu^4)\big|_{L=J/\Delta_L}\notag.
\end{align}
Setting $d=3$ and $\alpha=0$ in the above formula we reproduce \eqref{eq:largespin4d}.

Note that in heterotic string theory, the Gauss-Bonnet term $\alpha$ is positive \cite{Boulware:1985wk}, so we see that for $L\gg1$ the anomalous dimensions $\gamma_i$ increase in magnitude when the Gauss-Bonnet term is present. It would be interesting to understand whether this pattern persists for more general higher derivative interactions arising from string theory.

\subsection{Orbits in $dS_4$}

It is also curious to consider stable orbits in de Sitter. These exist only in $dS_4$ and the corresponding black hole metric can be obtained by changing $R_{AdS} \to i R_{dS}$. Denoting $\mu \equiv {G M \over R_{dS}}$ the black hole solution only exists for
\be
\mu \leq {2 \over 3 \sqrt{3}},
\ee
where $\mu = {2 \over 3 \sqrt{3}}$ corresponds to the extremal Nariai solution. 

Let us discuss circular orbits in this case. Using $V'(r_0)=0$ and setting $R_{dS}=1$, the conserved energy and angular momentum take the form
\begin{align}
\label{eq:orbitsQNdS}
E&=\frac{f(r_0)}{\sqrt{1-3 \mu/(2r_0)}}, ~~~~~ L=r_0^2 \frac{\sqrt{\mu/(2 r_0^3) - 1}}{\sqrt{1-3 \mu/(2r_0)}} \ .
\end{align}
From the structure of the square roots, time-like orbits only exist for
\be
\label{eq:orbitsdS}
{3 \over 2} \mu \leq r \leq \left( {\mu \over 2} \right)^{1/3} . 
\ee 
Moreover, imposing stability of the orbits, $V''(r_0)>0$ leads to the following constraint,
\be
(r-3 \mu) - 8 r^3 \left({r \over \mu} - {15 \over 8}\right)>0, ~~~\mu < {4 \over 75 \sqrt{3}} .
\ee
In particular, for small $\mu \ll 1$, the stability condition takes the form
\begin{align}
 3\mu<r<\frac{\mu^{1/3}}{2} .
\end{align}
In sharp contrast to the AdS case, stable orbits have spin which is bounded both from below and from above. The existence of a maximal spin for Regge trajectories in dS was also discussed in \cite{Noumi:2019ohm,Lust:2019lmq}.

\section{Decay of orbits}
\label{sec:raddecay}

We now turn to the decay of the orbits. The two relevant decay processes are gravitational radiation and tunneling into the black hole. The essential difference between the two is that gravitational radiation is ${1 \over c_T}$ suppressed and therefore is suppressed in the $c_T \to \infty$ limit. 

We will first compute the decay rate due to  radiation, and then compare the answer with the tunneling rate. Rather than deal with gravitational perturbations directly, we will consider the simpler case of scalar radiation. This model is defined by coupling a massless scalar field $\Phi$ to the orbit, 
\begin{align}
S=-\frac{1}{2}\int d^4 x\, \sqrt{-g}g^{\mu\nu}\partial_\mu \Phi\partial_\nu \Phi-\kappa \int d\tau \, (-g_{\mu\nu}\dot{x}^\mu(\tau)\dot{x}^\nu(\tau))^{1/2}\Phi.
\end{align}
 We work in $d=3$ for simplicity. In higher dimensions the functional form of the decay rate will be different, but the scaling with $\kappa$ is the same  as in $d=3$.  Also, in this section we will strictly consider the case of large black holes $\mu\gg1$, for which WKB methods are able to capture the leading contribution to the radiation.

\subsection{Radiation from circular orbits}\label{radcircular}

We are interested in computing the flux of energy and angular momentum through the horizon from the radiation field. These take the form 
\begin{align}
\frac{dE}{dt}&=-\int_{r=r_s} d^2 \Omega \, r^2 \partial_t \Phi \partial_z\Phi\label{energyexp}\\
\frac{dL}{dt}&=-\int_{r=r_s}d^2 \Omega\, r^2 \partial_\phi \Phi \partial_z \Phi\label{momentumexp}.
\end{align}
Here we have introduced the tortoise coordinate $z$ defined as follows,
\be
z = \int_r^{\infty} {d r' \over f(r')} .\label{tortoise}
\ee
In particular, $d z = -{dr \over f(r)}$. The black hole exterior corresponds to $z \in (0, \infty)$, where $z \to 0$ corresponds to the AdS boundary, and $z \to \infty$ being the black hole horizon. In this section we will consider circular orbits, leaving the general non-circular case to Appendix \ref{radnoncircular}.\\
\indent Let us first review the setup. We would like to solve the wave equation for a circular source at $r=r_0$. For $r_0\sim \mu $, we will see that the tunneling rate dominates over the radiation, so it suffices to analyze the radiation for $r_0\gg \mu$. Separating into Fourier modes,
\begin{align}
\Phi(t,z,\theta,\phi)=\sum_{J m}\int d\omega \,\frac{1}{r(z)}e^{-i\omega t}Y_{J m}(\theta,\phi)\psi_{Jm\omega}(z).\label{modeexpansion}
\end{align}
 The wave equation becomes \cite{Cardoso:2002up}
\begin{align}
\psi_{Jm\omega}''(z)+(\omega^2-V(r(z)))\psi_{Jm\omega}(z)= j_{Jm}(z), 
\end{align}
where 
\begin{align}
j_{Jm}(z)&=\kappa \int d\tau\, dt\, d\theta\, d\phi \, r(z)e^{i\omega t-im\phi}Y_{Jm}^*(\theta,0)\delta(t-\tau)\delta(r(z)-r_0)\delta(\phi-\tau)\delta(\theta-\pi/2) \notag\\
&=\kappa r_0 Y_{J m}^*(\pi/2,0)\delta(\omega-m)\delta(r(z)-r_0).\label{sourcej}
\end{align}
In deriving this equation we have used $r_0\gg \mu\gg 1$. The potential $V$ is given by 
\begin{align}
V(r)=f(r)\left(\frac{J(J+1)}{r^2}+2+\frac{\mu}{r^3}\right).\label{potentialfull}
\end{align}
This potential is applicable for massless radiation, and is therefore different from the orbital potential (\ref{eq:potentialBH}), which describes the classical motion of heavy particles.\\
\indent The general solution to the radial part of the equation is then \cite{Arfken:379118}
\begin{align}\label{gensolution}
\psi_{Jm\omega}(z)&=c_1\psi^1_{J\omega}(z)+c_2\psi^2_{J\omega}(z)\\
&\hspace{10 mm}+\frac{1}{W_{J\omega}}\int_{r(z)}^{\infty}\frac{dr'}{f(r')}\, (\psi^1_{J\omega}(z)\psi^2_{J\omega}(z'(r'))-\psi^2_{J\omega}(z)\psi^1_{J\omega}(z'(r')))j_{Jm}(z'(r'))\notag.
\end{align} 
Here $\psi^1$ and $\psi^2$ are solutions to the wave equation without the source, satisfying the boundary conditions
\begin{align}
\psi^1_{J\omega}(z)&\sim e^{-i\omega z}\text{ as }z\to \infty,\\
\psi^1_{J\omega}(z)&\sim \frac{A_{J\omega}}{z}+B_{J\omega}z^{2}\text{ as }z\to 0,\label{bcspsi1}\\
\psi^2_{J\omega}(z)&\sim z^{2}\text{ as }z\to 0,\label{bcs}\\
\psi^2_{J\omega}(z)&\sim C_{J\omega}e^{i\omega z}+D_{J\omega}e^{-i\omega z} \text{ as }z\to \infty.
\end{align}
In other words, $\psi^2$ is a normalizable mode, and $\psi^1$ satisfies purely ingoing boundary conditions. The Wronskian $W$, which is independent of $z$, is given by
\begin{align}
W_{J\omega}\equiv\psi^1_{J\omega}\partial_z \psi^2_{J\omega}-\psi^2_{J\omega}\partial_z \psi^1_{J\omega}=2iC_{J\omega} \omega=3A_{J\omega}.\label{wronskian}
\end{align}
\indent We compute the integration constants $c_1$ and $c_2$ as follows. We need purely ingoing boundary conditions at the horizon, and we need $\psi\to 0$ as $z\to0$ since we are interested in a normalizable solution. Near the horizon, the term proportional to $\psi^2$ must cancel because it involves an outgoing mode.  Therefore 
\begin{align}
c_2&=\frac{1}{W_{J\omega}} \int_{r_s}^{\infty}\frac{dr'}{f(r')}\, \psi^1_{J\omega}(z'(r'))j_{Jm}(z'(r'))
\end{align}
Near the boundary, the integral in (\ref{gensolution}) is zero, since the source is supported at $r'=r_0$. From (\ref{bcspsi1}) we have $\psi^1(z)\sim A/z$ for $z\sim 0$, so in order for $\psi$ to vanish at the boundary we need $c_1=0$. \\
\indent Plugging these values of $c_1$ and $c_2$ into (\ref{gensolution}), we now look at the behavior of $\psi$ near the horizon $z\to\infty$. This gives
\begin{align}
\psi_{Jm\omega }(z)&\sim \frac{\psi^1_{J\omega}(z)}{W_{J\omega}}\int_{r_s}^{\infty}\frac{dr'}{f(r')}\psi^2_{J\omega}(z'(r'))j_{Jm}(z'(r'))\notag\\
&\sim \frac{\kappa r_0Y_{Jm}^*(\pi/2,0)\psi^2_{J\omega}(z(r_0))}{2iC_{J\omega}\omega f(r_0)}\delta(\omega-m)e^{-i\omega z},\hspace{10 mm}z\to\infty\label{psinearrs},
\end{align}
where in the second line we used (\ref{sourcej}) and (\ref{wronskian}). From (\ref{bcs}), we see $\psi^2(z(r_0))$ behaves as $1/r_0^2$ as $r_0$ goes to infinity. Plugging (\ref{psinearrs}) into (\ref{modeexpansion}) and taking $r_0\to \infty$, we find 
\begin{align}
\Phi(t,z,\theta,\phi)&\sim \sum_{J m}\frac{\kappa }{2iC_{Jm} mr_sr_0^3}Y_{J m}(\pi/2,0)^*Y_{J m}(\theta,\phi)e^{-im(t+z)},\hspace{10 mm}z\to \infty.
\end{align}\\ 
Using (\ref{energyexp}) and (\ref{momentumexp}), the radiated power and angular momentum is (neglecting an order one constant)
\begin{align}
\left(\frac{dL}{dt}\right)_{J m}=\left(\frac{dE}{dt}\right)_{J m}\propto\frac{\kappa^2 |Y_{J m}(\pi/2,0)|^2}{|C_{Jm}|^2r_0^6}.\label{radiatedpower}
\end{align}
The equality of the rate of energy loss and angular momentum loss is a consequence of the delta function at $\omega=m$ in the source (\ref{sourcej}). Our task is now to solve for the coefficient $C_{Jm}$.\\
\indent In the limit $J\to \infty$, we can use a WKB analysis to solve the wave equation \cite{Festuccia:2008zx}. Luckily, it turns out that the power spectrum is dominated by this regime. This is completely different from stable orbits in flat space, where the dominant frequency is equal to the orbital frequency, and higher harmonics are suppressed. At large $J$ the potential (\ref{potentialfull}) takes the form 
\begin{align}
V(r)=\frac{f(r)}{r^2}J(J+1).
\end{align}
With the WKB ansatz $\psi^2=e^{J S}$, the wave equation becomes
\begin{align}
(\partial_z S)^2=\frac{J(J+1)-m^2}{J^2}+\frac{1}{r^2}-\frac{\mu }{r^3}.\label{WKBaction}
\end{align}
There is a single turning point $r_t$ outside of the horizon. The region $r<r_t$ is the classically allowed region of a particle in the potential. For $r$ outside of this turning point, the wavefunction is exponentially decaying,
\begin{align}
\psi^2_{Jm}(z)\propto \frac{1}{(\partial_z S)^{1/2}}\exp\left(J\int_{r(z)}^{\infty}\frac{dr'}{f(r')}\partial_z S\right).
\end{align}
\indent Now we need to fix the proper normalization. The WKB expansion fails near the boundary, where the potential (\ref{potentialfull}) takes the form 
\begin{align}
V(r)=2r^2+J(J+1).
\end{align}
The solution to the wave equation near the boundary is therefore
\begin{align}
\psi^2_{J m}(z)=\frac{r(z)^{-1/2}J_{3/2}(\sqrt{m^2-J(J+1)}/r(z))}{(J(J+1)-m^2)^{3/4}}.
\end{align}
We have chosen the normalization so that (\ref{bcs}) is satisfied. Expanding for $J/r\gg 1$, we get 
\begin{align}
\psi^2_{Jm}(z)\sim \frac{\exp(\sqrt{J(J+1)-m^2}/r(z))}{J(J+1)-m^2}.
\end{align}
This fixes the normalization of the WKB wavefunction, 
\begin{align}
\psi^2_{Jm}(z)=\frac{1}{\sqrt{J}(J(J+1)-m^2)^{3/4}} \frac{1}{(\partial_z S)^{1/2}}\exp\left(J\int_{r(z)}^{\infty}\frac{dr'}{f(r')}\partial_z S\right).\label{wvfnc}
\end{align}
\indent The integral (\ref{wvfnc}) can be done to examine the behavior of the wavefunction near the turning point. Let us start with the case $J\ll \mu^2$. Then we can drop the $1/r^2$ term in (\ref{WKBaction}), since the first term in (\ref{WKBaction}) is bounded below by $1/J$. In this case the turning point is at
\begin{align}
r_t=\left(\frac{J^2 \mu }{J(J+1)-m^2}\right)^{1/3},\hspace{5 mm}J\ll \mu^2.
\end{align}
At the turning point, the exponent in (\ref{wvfnc}) becomes \begin{align}
J\int_{r_t}^{\infty}\frac{dr}{f(r)}\partial_z S=\sqrt{J(J+1)-m^2}\int_{r_t}^{\infty}\frac{dr}{r^2-\mu/r}\sqrt{1-r_t^3/r^3}.\label{exponentsmallspin}
  \end{align}
There are two limits we can take in (\ref{exponentsmallspin}). The first is $r_t\to \mu^{1/3}$, or $|m|\ll J$. We get 
\begin{align}
J\int_{r_t}^{\infty}\frac{dr}{f(r)}\partial_z S\sim \frac{J}{\mu^{1/3}},\hspace{10 mm}|m|\ll J\ll \mu^2.
\end{align}
The second limit is $r_t\to \infty$, or $|m|-J\ll J$. Then (\ref{exponentsmallspin}) becomes
\begin{align}
J\int_{r_t}^{\infty}\frac{dr}{f(r)}\partial_z S\sim \frac{c(J(J+1)-m^2)^{5/6}}{(J^2\mu)^{1/3}},\hspace{10 mm}|m|-J\ll J\ll \mu^2,\label{wkbsdominant}
\end{align}
 where $c= \sqrt{\pi}\Gamma(4/3)/(2\Gamma(11/6))$. \\
\indent We now turn to the opposite case of large spin $J\gg \mu^2$. In this limit the turning point approaches $r=\mu$, and the exponent in (\ref{wvfnc}) is 
\begin{align}
J\int_{\mu}^{\infty}\frac{dr}{r^2}\sqrt{\frac{1}{r^2}-\frac{\mu}{r^3}}=\frac{4J}{15\mu^2},\hspace{10 mm}J\gg \mu^2\label{cutoff}.
\end{align}
\indent We can now use the standard WKB connection formulas to write down the oscillating solution in the classically allowed region. But actually we just need the magnitude of $C_{Jm}$, which can be computed by evaluating the magnitude of the solution (\ref{wvfnc}) at the horizon. The dominant region is $J\ll \mu^2$. In this limit we find from (\ref{wkbsdominant})
\begin{align}
|C_{Jm}|=\frac{1}{\sqrt{J}(J(J+1)-m^2)^{3/4}}\exp\left(\frac{2c(J(J+1)-m^2)^{5/6}}{(J^2\mu)^{1/3}}\right), 
\end{align}
so the radiated power (\ref{radiatedpower}) is
\begin{align}
P_{J}=\frac{\kappa^2}{r_0^6}\sum_{m=-J}^{J}|Y_{J m}(\pi/2,0)|^2J(J(J+1)-m^2)^{3/2}\exp\left(-\frac{2c(J(J+1)-m^2)^{5/6}}{(J^2\mu)^{1/3}}\right).
\end{align}
Because of the exponential factor, the sum is dominated by $1-|m|/J\ll 1$. In this limit the spherical harmonics take the form 
\begin{align}
|Y_{J m}(\pi/2,0)|^2\sim J^{1/4}\frac{\Gamma(2n+1)}{2^{2n}\Gamma(n+1)^2}.
\end{align}
where we have defined $J=|m|+2n$. We finally get
\begin{align}
P_{J}&\sim \frac{\kappa^2J^3 }{r_0^6}\sum_{n=0}^{\infty}\frac{\Gamma(2n+1)}{2^{2n}\Gamma(n+1)^2}(1+4n)^{3/2}\exp\left(-\frac{2c(1+4n)^{5/6}J^{1/6}}{\mu^{1/3}}\right)\notag\\
&\sim \frac{\kappa^2J^3 }{r_0^6}\left(\frac{\mu^2}{J}\right)^{2/5}.\label{dominant}
\end{align}
 \begin{figure}[t]
\includegraphics[scale=.65]{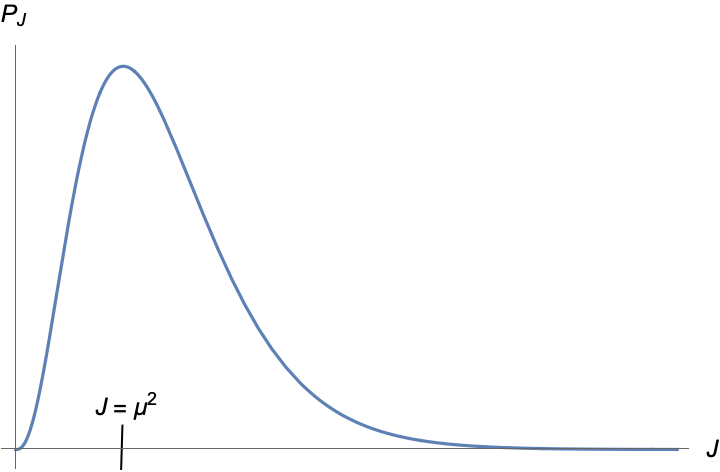}
\centering
\caption{The power spectrum $P_J$ grows like a power law until $J\sim \mu^2$, at which point it begins to decay exponentially.}
\centering
\end{figure} 
\indent In order to compute the total power, we need to sum (\ref{dominant}) over all $J$. This sum is divergent, but there is an effective cutoff at $J=\mu^2$. Indeed, from (\ref{cutoff}) we find that for $J\gg \mu^2$
\begin{align}\label{powerlargeJ}
P_J\propto \frac{1}{|C_{Jm}|^2}\sim \exp\left(-\frac{8J}{15\mu^2}\right),\hspace{10 mm}J\gg \mu^2.
\end{align}
We can therefore approximate the sum over $J$ by only summing up to $J=\mu^2$, and the total power is
\begin{align}
\sum_{J}P_{J}&\approx \sum_{J<\mu^2} \frac{\kappa^2J^3 }{r_0^6}\left(\frac{\mu^2}{J}\right)^{2/5}\notag\\
&\sim \frac{\kappa ^2 \mu^8}{r_0^6}.\label{totalpower}
\end{align}
 A similar analysis gives the radiation power for a general massive scalar field with dimension $\Delta$,
 \begin{align}
 \sum_{J<\mu^2}P_{J}\sim \frac{\kappa ^2 \mu^{2\Delta+2}}{r_0^{2\Delta}},\
 \end{align}
 At large $r_0$, the lowest dimension operators dominate the radiation spectrum. \\
 \indent To summarize, the result of the analysis is a power spectrum that is peaked at very large $J\sim \mu^2$. Since $|m|$ is close to $J$, the radiation is emitted near the equatorial plane. Note that for black holes near the Hawking-Page transition, the $|m|=J$ mode is dominant over the others. This is purely synchrotronic radiation, as found numerically in \cite{Cardoso:2002up,Brito:2021qiw}. In our case we are considering black holes with $\mu\gg 1$, so the radiation is not purely synchrotronic. 
 
\subsection{Decay of circular orbits}

Now that we have found the emitted power, we can compute the decay time. For the purposes of this section, we will only consider the case where the energy of the orbit is much smaller than the mass of the black hole. The energy of a circular orbit with $r_0\gg \mu$ and $\Delta_L\gg 1$ is $
\Delta_Lr_0^2$, so we require $r_0^2\ll M/\Delta_L$, along with $r_0\gtrsim \mu$ for stability. These two conditions can only be satisfied above the Hawking-Page transition if 
\begin{align}
\label{eq:regionofM}
\frac{1}{G}<M\ll \frac{1}{\Delta_L G^2}.
\end{align}
In terms of the temperature, we have 
\begin{align}
1< T\ll \frac{1}{(\Delta_L G)^{1/3}}.
\end{align} 
In this intermediate range of temperatures, the final state after decay is a slowly spinning black hole whose mass is slightly larger than the original black hole, and we can use the results of Section \ref{radcircular} to analyze the decay process. \\
\indent Since the decay process is adiabatic, the orbit remains approximately circular with radius $r_0(t)$ when radiation is taken into account. The energy of the orbit is $\Delta_L r_0(t)^2$ for $r_0\gg \mu$, so the rate of energy loss is
\begin{align}
\frac{d(\Delta_Lr_0^2)}{dt}=2\Delta_L r_0\dot{r_0}. 
\end{align} 
Equating this to the radiated power (\ref{totalpower}) gives 
\begin{align}
\dot{r}_0\sim- \kappa^2\frac{\mu^8}{\Delta_L r_0^7},
\end{align}
so it follows that the decay time is 
\begin{align}\label{decaytimescalar}
t_{\text{radiation}}&\sim \frac{\Delta_L}{\kappa^2 }\left(\frac{r_0}{\mu}\right)^8.
\end{align}
This goes to infinity at large $r_0$, where the orbit approaches the asymptotically AdS region.\\
\indent Let us briefly comment on the case of gravitational radiation. Naively, the power emitted through gravity waves can be computed by replacing $\kappa$ with $\Delta_L\sqrt{1/c_T}$ \cite{Misner:1972jf}, and the corresponding decay time is
\begin{align}
t_{\text{radiation}}\sim \frac{c_T}{\Delta_L^5}\frac{J^4}{\mu^8}.\label{decaytime}
\end{align}
However, there are subtleties involved in replacing $\kappa$ with $\Delta_L\sqrt{1/c_T}$ when computing the radiation from an unstable orbit around an asymptotically flat black hole \cite{Breuer:1973kt,Chitre:1972fv,Davis:1972dm}. It would be interesting to explicitly compute the gravitational radiation spectrum and verify whether (\ref{decaytime}) holds.

\subsection{Radiation vs. tunneling}\label{radvstunn}
\indent Let us now compare the relative effects of radiation and tunneling. Since the potential barrier is finite, there is a small probability for the particle to tunnel over the barrier into the black hole horizon. This is the sole source of instability at infinite $c_T$, where gravitational radiation can be neglected. The tunneling rate is exponentially suppressed in $\Delta_L$, since the action for a point particle is proportional to $\Delta_L$. Therefore tunneling is not captured by the Bohr-Sommerfeld approximation (\ref{eq:BSclassical}). The tunneling rate is related to the imaginary part of the quasi-normal mode energy, \cite{Festuccia:2008zx}
\begin{align}
\Gamma=|\text{Im }\omega_n|=\exp\left(-2\Delta_L \int_{z_b}^{z_c} dz\, \sqrt{V(z)-E_n^2}\right).
\end{align}
Here $z_c$ is the third turning point, which is over the potential barrier (see Figure \ref{potentialfig}) . \\
\indent The decay rate simplifies in the limiting case of a circular orbit that approaches the boundary. In this limit we have $E_n\sim L$, $r(z_c)\sim \mu^{1/(d-2)}$, and $r(z_b)\sim \sqrt{L}\to \infty$. The tunneling rate becomes 
\begin{align}\label{tunnelinggend}
\Gamma=\exp\left(-2J \int_{\mu^{1/(d-2)}}^{\infty}\frac{dr}{rf(r)}\sqrt{1-\frac{\mu}{r^{d-2}}}\right) .
\end{align}
For large black holes $\mu\gg1$, we have $\mu^{1/(d-2)}\gg r_s$, so we can replace $f(r)$ by $r^2$ in the denominator of (\ref{tunnelinggend}. We can then perform the integral to find
\begin{align}
\Gamma=\exp\left(-\frac{J}{\mu^{2/(d-2)}}\frac{\sqrt{\pi}\Gamma\left(\frac{d}{d-2}\right)}{2\Gamma\left(\frac{d}{d-2}+\frac{1}{2}\right)}\right),\hspace{10 mm}\mu\gg 1,\frac{J}{\mu^{2/(d-2)}}\gg 1.\label{tunnlargemu}
  \end{align}
  For small black holes with $\mu\ll 1$, we consider the region of the integral where $r/ \mu^{1/(d-2)}$ is held fixed as $\mu\to0$. Then we can replace $f(r)$ by $1-\mu/r^{d-2}$ in the denominator of (\ref{tunnelinggend}), finding
  \begin{align}\label{tunnsmallmu}
      \Gamma&=\exp\left(-2J \int_{\mu^{1/(d-2)}}\frac{dr}{r\sqrt{1-\frac{\mu}{r^{d-2}}}}\right)\approx\mu^{\frac{2J}{d-2}},\hspace{10 mm}\mu\ll 1,J\gg 1.
  \end{align}
  In $d=3$ this reproduces the leading behavior at large $J$ of the imaginary part of the quasi-normal mode energy of a small black hole, see eq. (112) in \cite{berti}.\\
 \indent We now compare the tunneling rate to the radiation rate for large black holes in $d=3$. From (\ref{tunnlargemu}), we find that the characteristic time scale for tunneling is
\begin{align}
t_{\text{tunneling}}=\exp\left(\frac{8J}{15\mu^2}\right).
\end{align}
Therefore tunneling occurs before the radiative decay time (\ref{decaytimescalar}) in $d=3$ if
\begin{align}
\exp\left(\frac{8J}{15\mu^2}\right)\ll\frac{1}{\kappa^2 \Delta_L^3}\frac{J^4}{\mu^8}.\label{inequality}
\end{align}
Replacing $\kappa$ by $\Delta_L\sqrt{1/c_T}$ as is appropriate for gravitational radiation, the equation (\ref{inequality}) requires 
\begin{align}
\Delta_L\mu^2< J\ll \mu^2\log  c_T.
\end{align}
For larger $r_0$, the radiation begins to dominate. In this regime the imaginary part of the quasi-normal mode energy is controlled by $1/c_T$ in the presence of gravitational radiation. It would be interesting to verify this by analyzing the poles of the retarded Green function.

\section{Orbit states and the light-cone bootstrap}
\label{sec:orbitsLCbootstrap}

In this section we consider the four-point function of scalar primary operators in a theory with a classical gravity dual.  In terms of the CFT data we assume that the CFT central charge (or equivalently, the two-point function of the stress-energy tensor) is very large, $c_T \gg 1$. We also assume that the gap in the spectrum of single-trace higher spin operators is large, $\Delta_{\text{gap}} \gg 1$. 

We take a pair of operators to be heavy, $\Delta_H \sim c_T$, and think of them as creating a classical black hole background as we take $c_T \to \infty$. The second pair of operators we take to be light, $\Delta_L \ll c_T$, and we will use them to probe the background created by the heavy operators. This is the setup considered recently in \cite{\citeHLLH}.

Let us start by setting up some basic conventions (we closely follow the conventions of \cite{Jafferis:2017zna}). We define the four-point function as follows
\be
\label{eq:basicdef}
G(z,\bar z) \equiv \langle \cO_H(0) \cO_L(z,\bar z)  \cO_L(1,1) \cO_H(\infty) \rangle ,
\ee
where all operators for simplicity are taken to be real scalars. As usual $\cO_H(\infty) = \lim_{x_4 \to \infty} |x_4|^{\Delta_H} \cO_H(x_4)$. In writing the formula above we used conformal symmetry to put all four operators in a two-dimensional plane, on which we define the coordinate $z = x^1 + i x^2$. As usual in Euclidean space $\bar z = z^*$, and upon Wick rotation $z$ and $\bar z$ become real and independent.

The OPE expansion in different channels takes the form
\be
\label{eq:schannel}
s\text{-channel}: ~~~G(z,\bar z) &= (z \bar z)^{- {1 \over 2} (\Delta_H + \Delta_{L}) } \\
&\hspace{5 mm}\times \sum_{{\cal O}_{\Delta, J}} | \lambda_{H,L, {\cal O}_{\Delta, J}} |^2 g_{\Delta, J}^{\Delta_{H,L}, - \Delta_{H,L}}(z, \bar z), ~~~ |z|<1 \notag \ .
\ee
\be
\label{eq:tchannel}
t\text{-channel}: ~~~G(z,\bar z) &= \left([1-z][1- \bar z]\right)^{- \Delta_L } \\
&\hspace{5 mm}\times\sum_{ {\cal O}_{\Delta, J} } \lambda_{L,L, {\cal O}_{\Delta, J}} \lambda_{H,H, {\cal O}_{\Delta, J}} g_{\Delta, J}^{0,0}(1-z,1-\bar z) ,\quad   |1-z|<1\notag \ .
\ee
\be
\label{eq:uchannel}
u\text{-channel}: ~~~G(z,\bar z) &= (z \bar z)^{{1 \over 2} (\Delta_H - \Delta_{L}) } \\
&\hspace{5 mm}\times\sum_{ {\cal O}_{\Delta, J} } | \lambda_{H,L, {\cal O}_{\Delta, J}} |^2 g_{\Delta, J}^{\Delta_{H,L}, - \Delta_{H,L}}\left({1 \over z}, {1 \over \bar z} \right),
~~~ |z|>1,\notag \ 
\ee
where we have defined $\Delta_{H,L} = \Delta_{H} - \Delta_{L}$, and the sum is over an infinite set of primary operators labeled by their scaling dimension $\Delta$ and spin $J$. 
 
It is also convenient to define
\be
g(z,\bar z) \equiv (z\bar z)^{{1\over 2} \Delta_L}  G(z,\bar z) \ .
\ee
The correlation function is invariant under the exchange of the locations of the two light operators. This is encoded in the crossing equation in the $s$ and $u$ channels,
\be
g(z, \bar z) =  g \left({1 \over z}, {1 \over \bar z} \right) \ .
\ee
In the case where the operators are charged, the corresponding formulas can be found in \cite{Jafferis:2017zna}.

The formulas above are completely general. We would like to focus on the situation where $c_T \to \infty$ with ${\Delta_H \over c_T}$ kept fixed. In this limit we think of $\Delta_H$ as creating a classical black hole background.

Let us consider the $s$- and $u$-channel OPEs. As explained in \cite{Jafferis:2017zna}, the conformal blocks simplify in this limit because descendants are suppressed by the factor
\be
\label{eq:conditiondesc}
{(\Delta - \Delta_H)^2 \over \Delta_H } \ll 1 .
\ee
At this point it is not obvious that in the limit of interest the relevant primary operators that appear in the OPE satisfy \eqref{eq:conditiondesc}, but later using the eigenstate thermalization hypothesis (ETH) we will check that this is indeed the case.

Therefore the relevant conformal blocks are simply
\be
 g_{\Delta, J}^{\Delta_{H,L}, - \Delta_{H,L}}(z, \bar z) = (z \bar z)^{{\Delta \over 2}} P_J^{(d)} \Big( {z + \bar z \over 2 \sqrt{z \bar z}} \Big) + O \Big({(\Delta - \Delta_H)^2 \over \Delta_H} \Big), 
\ee
where 
\begin{align}
P_J^{(d)}(x) \equiv {\Gamma({d-2 \over 2}) \Gamma(J+1) \over \Gamma(J+{d \over 2}-1)} C_J^{({d \over 2}-1)}(x),
\end{align} 
and $C_J^{(\alpha)}(x)$ are the standard Gegenbauer polynomials. For $d=3$ these are the usual Legendre polynomials.

\subsection{OPE and ETH}

We would next like to rewrite the expansion above using the eigenstate thermalization hypothesis (ETH) \cite{srednicki1999approach}, see \cite{DAlessio:2015qtq} for review, which states that \cite{Lashkari:2016vgj,Delacretaz:2020nit}
\be
\label{eq:ETH}
\langle E_H | {\cal O}_L | E_{H'} \rangle = {\cal O}_L(E_H) \delta_{H, H'} + e^{-{1\over 2} S(\b E)} f_{{\cal O}_L}(\b E, \omega) R_{H H'},
\ee
where $\b E = {1\over 2}(E_H + E_{H'})$, $\omega = E_{H'} - E_H$. The function ${\cal O}_L(E_H)$ is a smooth function of the energy given by the microcanonical average of ${\cal O}_L$. The matrix $R_{H H'}$ is a random matrix with zero mean and unit variance (this is not exactly correct \cite{Foini:2018sdb} but it will be sufficient for our purposes).\footnote{See also a related discussion in \cite{Dymarsky:2018ccu}.} The function $f_{{\cal O}_L}(E, \omega)$ is a smooth function of both variables closely related to the thermal two-point function. 

In the formula \eqref{eq:ETH} above we used energy eigenstates of the theory on the cylinder. We would like to switch to the plane
and use the operator-state correspondence. The mapping takes the form
\be
ds^2_{\text{cyl}} = d \tau^2 + R^2 d \Omega_{d-1}^2 = \Big( {R \over r} \Big)^2 \Big( d r^2 + r^2 d \Omega_{d-1}^2  \Big), ~~~ \tau = R \log r . 
\ee
Under this conformal map we have
\be
\langle H | \cO(\tau, \vec n) | H' \rangle_{\text{cyl}} = \Big( {r \over R} \Big)^{\Delta_\cO } { \la {\cal O}_{H'}(0) {\cal O}(x) {\cal O}_H^\dagger (\infty) \ra_{{\mathbb R}^d} \over \sqrt{\la {\cal O}_{H'}(0)  {\cal O}_{H'}^\dagger (\infty) \ra_{{\mathbb R}^d} \la {\cal O}_{H}(0) {\cal O}_H^\dagger (\infty) \ra_{{\mathbb R}^d}} } ,
\ee
where $r = e^{{\tau \over R}}$ and $\vec x = r \vec n$. In particular, the energy on the cylinder is related to the scaling dimension by $E_H = {\Delta_H \over R}$. Via this mapping the ETH ansatz \eqref{eq:ETH} becomes a statement about the three-point functions that appear in the heavy-light OPE channel. For example, setting $H' = H$ we get
\be
\lambda_{H,H,\cO_{\Delta,J}} = { \la {\cal O}_{H}(0) {\cal O}(1) {\cal O}_H^\dagger (\infty) \ra_{{\mathbb R}^d} \over \la {\cal O}_{H}(0)  {\cal O}_{H}^\dagger (\infty) \ra_{{\mathbb R}^d} } = R^{\Delta_\cO} \la H | \cO_{\Delta,J}(0) | H \ra_{\text{cyl}}.
\ee

Using the ETH ansatz we find for the four-point function in the $s$-channel (after trivially averaging over $R_{HH'}$)
\be
&g(z, \b z) \equiv (z \b z)^{{1\over 2} \Delta_L}G\left( z , \b z  \right) = R^{2\Delta_L} \Bigg[ {\cal O}_L(\Delta_H)^2 \notag\\
& + \sum_{\Delta_{H'}, J}  e^{-S(\b \Delta/R)} \left|f_J\left({\b \Delta \over R}, {\omega \over R}\right) \right|^2 (z \b z)^{{1\over 2} (\Delta_{H'} - \Delta_H) } P_J^{(d)} \left( z + \b z \over 2 \sqrt{z \b z} \right) \Bigg], \notag\\
&\b \Delta = {\Delta_H + \Delta_{H'} \over 2}, \qquad \omega = \Delta_{H'} - \Delta_H ,
\ee
where we used that descendants are suppressed by powers of $\Delta_H$. The sum is over primary operators. To simplify the formulas and avoid extra clutter we next set $R=1$.

Assuming an approximately continuous spectrum and introducing the corresponding density of states we convert the OPE sum into an integral,\footnote{In principle, we should write at this point $S_J(\Delta_H)$. However, using the symmetry of the Kerr black holes under $J \to - J$, we immediately see that $S_J(\Delta_H) - S(\Delta_H) \sim {1 \over c_T}$ for $\Delta_H \sim c_T \to \infty$ and $J \sim O(1)$. It would be interesting to understand if this is in fact true in any large $c_T$ CFT.}
\be
\label{eq:sumtoint}
\sum_{\Delta_{H'}} \to \int d \Delta_{H'}\,  e^{S(\Delta_{H'})} = \int_{-\infty}^\infty d \omega ~e^{S\left( \Delta_H + \omega \right)} .
\ee
Using that $z =e^{\tau+ i \theta}$ and $\overline{z}=e^{\tau - i \theta}$, we have in the $s,u$-channels for the connected correlator
\be
\label{eq:gcon}
g_c(z,\b z) &\equiv g(z,\b z) -{\cal O}_L(\Delta_H)^2 \\
& = \sum_{J = 0}^\infty  \int_{-\infty}^\infty d \omega 
~ e^{ S(\Delta_H + \omega) - S\left( \Delta_H + {\omega \over 2} \right) } e^{- |\tau| \omega } \left|f_J\left(\Delta_H + {\omega \over 2} , \omega \right) \right|^2 P_J^{(d)}\left( \cos \theta \right),\notag
\ee
where $\tau <0$ for the $s$-channel and $\tau >0$ in the $u$-channel. In the formula above we extended the integral over $\omega $ in \eqref{eq:sumtoint} and \eqref{eq:gcon} to $(-\infty, \infty)$, which makes sense in the $\Delta_H \to \infty$ limit. 

Next we would like to expand in the limit $\omega \ll \Delta_H$, as in \eqref{eq:conditiondesc}. We have
\be
\label{eq:entropyexp}
&S(\Delta_H + \omega) - S\left( \Delta_H + {\omega \over 2} \right) = {\omega \over 2} {\p S \over \p \Delta_H} + {3\over 8} \omega^2 {\p^2 S \over \p \Delta_H^2} + \dots \ ,\notag \\ 
&{\p S(\Delta_H) \over \p \Delta_H} = \beta , \qquad 
{\p^2 S(\Delta_H) \over \p \Delta_H^2} =- {1 \over T^2} {\p T \over \p E}  = - {\beta^2 \over C} \ , 
\ee
where $C = {\p E \over \p T}$ is the heat capacity. 
The expansion \eqref{eq:entropyexp} can be used when the second term is much smaller then the first term,
\be
\label{eq:expansionval}
\omega \ll {C \over \beta} \sim c_T \sim \Delta_H . 
\ee
Therefore the expansion is reliable as long as $\omega \ll \Delta_H$.

To summarize, the OPE expansion in the $s$- and $u$- channel takes the following form,
\be
\label{eq:EuclidOPE}
g(z, \bar z) =  \lambda_{H,H,L}^2 + \sum_{J = 0}^\infty  \int_{-\infty}^\infty d \omega\, e^{\left( {\beta \over 2} - |\tau| \right)\omega } \left|f_J\left(\Delta_H , \omega \right) \right|^2 P_J^{(d)}\left( \cos \theta \right) ,
\ee
where so far we used ETH to go from a sum over a continuum of heavy operators to an integral over a smooth function $\left|f_J\left(\Delta_H , \omega \right) \right|^2$.

Next let us connect the above representation to the thermal two-point function on the sphere. To this end we define the connected two-sided Wightman function
\be
\label{eq:two-sided}
\left\la \cO_L\left(t - i {\beta \over 2} , \theta\right) \cO_L(0,0) \right\ra_\beta = \sum_{J = 0}^\infty \int_{- \infty}^\infty d \omega\, e^{- i \omega t} g_J(\beta, \omega)  P_J^{(d)}(\cos \theta) . 
\ee 
Equivalently, we can take the Euclidean formula \eqref{eq:EuclidOPE} and continue it to Lorentzian time by setting $\tau = {\beta \over 2} + i t$.

The statement of the ETH hypothesis then leads to the following identification
\be
\label{eq:ETHrelation}
{\rm ETH}:~~~ \left|f_J\left(\Delta_H , \omega \right) \right|^2  = g_J(\beta, \omega) |_{\beta = \beta(\Delta_H)}.
\ee
The KMS condition (or equivalently, invariance of \eqref{eq:EuclidOPE} under $\tau \to \beta - \tau$, see e.g. \cite{Iliesiu:2018fao}) leads to
\be
\label{eq:KMSimage}
\text{KMS}: ~~~ g_J(\beta, \omega)  =g_J(\beta,-\omega).
\ee
Unitarity implies that $g_J(\beta, \omega)$ is real and non-negative for real $\omega$. This fact implies that away from the real axis it satisfies
\be
\label{eq:complexconjugation}
\text{Unitarity}: ~~~g_J(\beta, \omega^*) = \Big( g_J(\beta, \omega) \Big)^* . 
\ee

In deriving the formula above we assumed that $\omega \ll \sqrt{\Delta_H}$, see \eqref{eq:conditiondesc}: first, when neglecting the contribution of the descendants; second, when expanding the entropy. Let us check that this is indeed the case.

To this end recall the universal large $\omega$ asymptotic of $g_J(\beta, \omega)$,
\be\label{tauberian}
\lim_{\omega \to \infty} g_J(\beta, \omega) \sim \omega^{2 \Delta_L - d} e^{- {\beta \omega \over 2}} ,
\ee
which should be understood in the averaged/Tauberian sense \cite{Pappadopulo:2012jk}. It follows from reproducing correctly the $\tau \to 0$ behavior, which is controlled by the unit operator in the $t$-channel,
\be
g(e^{\tau},e^{\tau}) \sim {1 \over \tau^{2 \Delta_L}}, ~~~ \tau \to 0 .
\ee
More precisely, to derive \eqref{tauberian} one computes the spin $J$ projection of the contribution of the unit operator in the $t$-channel to $g_J(\beta, \omega)$. Taking the $\tau \to 0$ limit then gives \eqref{tauberian}. 

From \eqref{tauberian} we conclude that only operators with $\omega \sim {1 \over \beta} \sim O(1) \ll \sqrt{\Delta_H}$ contribute significantly to the OPE as has been assumed in the derivation above. 

\subsection{Quasi-normal modes}

Quasi-normal modes are defined as poles of $g_J(\beta, \omega)$, see e.g. \cite{Horowitz:1999jd,cardoso,Turiaci:2016cvo}. From the properties above it is clear that given a pole at $\omega_0$, $g_J(\beta, \omega)$ also has poles at $- \omega_0$, $\omega_0^*$, $- \omega_0^*$. The conclusion of this discussion is that QNMs manifest themselves as singularities of the averaged OPE coefficients continued in the complex plane as a function of $\omega = \Delta_{H'} - \Delta_{H}$. Below we will focus on QNMs with ${\rm Re}[\omega_{n,J}],{\rm Im}[\omega_{n,J}]>0$ with the residue $\lambda_{n,J}$, from which we obtain all other QNMs using the KMS symmetry \eqref{eq:KMSimage} and complex conjugation \eqref{eq:complexconjugation}.

In theories with a classical gravity dual it is known that poles are in fact the only singularities of $g_J(\beta, \omega)$. We can thus try to close the integration contour in \eqref{eq:EuclidOPE} to the upper half-plane to get an alternative expansion of the correlation function in terms of the QNMs. For convergence reasons we set
\be
\tau = \beta/2 - i t , 
\ee
which is naturally related to the two-sided thermal function $\left\la \cO_L\left(t - i \beta/2 , \theta\right) \cO_L(0,0) \right\ra_\beta$.

Introducing in this way
\be
g(t,\theta) \equiv  g \Big( e^{\beta/2 - i (t-\theta) }, e^{\beta/2 - i (t+\theta)} \Big) ,
\ee
we get by closing the $\omega$ integration contour in \eqref{eq:EuclidOPE} in the upper half-plane for $t>0$\footnote{For $t<0$ we can close the contour in the lower half-plane.}
\be
\label{eq:QNM}
g(t,\theta) =  \lambda_{H,H,L}^2 + 2 \pi i \sum_{n,J = 0}^\infty (\lambda_{n,J}  e^{i t \omega_{n,J} } - \lambda_{n,J}^*  e^{-i t \omega_{n,J}^* } )  P_J^{(d)}\left( \cos \theta \right) , ~~~ {\rm Re}[\omega_{n,J}],{\rm Im}[\omega_{n,J}]>0 ,
\ee
where we dropped the contribution of the arc at infinity thanks to the exponential suppression $e^{- t {\rm Im}[\omega]}$ of the integrand. In this way we get the following QNM representation of the two-sided correlator,
\be
\label{eq:QNMexpansion}
&\text{QNM}: ~~~g(t,\theta) =  \lambda_{H,H,L}^2 - 4 \pi \sum_{n,J = 0}^\infty {\rm Im} \lambda_{n,J}  e^{- |t|{\rm Im} \omega_{n,J} } \cos ( {\rm Re} \omega_{n,J} |t| )  P_J^{(d)}\left( \cos \theta \right)\nn \\
&-4 \pi \sum_{n,J = 0}^\infty {\rm Re} \lambda_{n,J}  e^{- |t|{\rm Im} \omega_{n,J} } \sin ({\rm Re} \omega_{n,J} |t| ) P_J^{(d)}\left( \cos \theta \right) , ~~~ {\rm Re}\omega_{n,J},{\rm Im}\omega_{n,J} >0 .
\ee
The imaginary part of ${\rm Im} \omega_{n,J} $ controls the decay rate in Lorentzian time. As in \cite{Jafferis:2017zna} we see that $g(t,\theta)$ is not manifestly analytic when expanded around $t=0$, which leads to an infinite set of sum rules which we do not explore in this paper.

In the present paper we assume that $\lambda_{H,H,L}=0$ to leading order in $c_T$ (or equivalently, that the thermal one-point function of the light operator is zero to leading order). In this case the hydrodynamic modes do not appear in the QNM expansion above, see \cite{Delacretaz:2020nit}.

Another useful way to think about QNMs is in terms of the conformal partial wave expansion of the four-point function \cite{Mack:2009mi,Costa:2012cb,Caron-Huot:2017vep},
\be
G(z,\bar z) = \sum_{J=0}^\infty \int_{{d \over 2} - i \infty}^{{d \over 2} + i \infty} {d \Delta \over 2 \pi i} c(\Delta,J) F_{\Delta,J}(z, \bar z).
\ee
The function $c(\Delta,J)$ is meromorphic and it contains poles at the position of operators with residues controlled by the three-point functions
\be
{\rm Res}_{\Delta = \Delta_H'} c(\Delta,J) \sim |\lambda_{H,L,H'}|^2 .
\ee
As we take the large $c_T$ limit these poles merge and form a cut. QNMs are then nothing but poles of $c(\Delta,J)$ on the second sheet!\footnote{In other words, they are somewhat analogous to resonances in S-matrix theory.}

\subsection{Relation to the light-cone bootstrap}

Let us next see how the unit operator in the $t$-channel is reproduced in more detail. It is useful first to write the precise formula in generalized free field theory. The heavy-light OPE expansion then takes the form
\be
\label{eq:unitoperatorGFF}
&\sum_{n,J=0}^{\infty} c_{n,J} e^{-(\Delta_L+2n+J)|\tau|}  P_J^{(d)}(\cos \theta) = {1 \over 2^{\Delta_L} (\cosh \tau - \cos \theta)^{\Delta_L}} , \nn \\
c_{n,J} &= \frac{\Gamma \left(\frac{d}{2}+J\right) \Gamma
   \left(-\frac{d}{2}+n+\Delta_L+1\right) \Gamma
   (J+n+\Delta_L)}{\Gamma (\Delta_L) \Gamma (J+1) \Gamma
   (n+1) \Gamma \left(-\frac{d}{2}+\Delta_L+1\right) \Gamma
   \left(\frac{d}{2}+J+n\right)} ,
\ee
 where $c_{n,J}$ correspond to the three-point functions of the double-twist operators with dimension $\Delta = \Delta_H + \Delta_L + 2n + J$ and spin $J$ in the limit $\Delta_H \to \infty$, see e.g \cite{Li:2019zba,Li:2020dqm}.  In agreement with \eqref{tauberian}, $c_{n,J} \sim n^{2 \Delta_L - d}$ at large $n$ and fixed $J$.

In an interacting theory the RHS represents the leading singularity in the light-cone limit and the LHS becomes more and more accurate at high spin (with computable corrections). In terms of the three-point functions discussed in the previous section, the result above takes the following form,
\be
\label{eq:discretesum}
e^{{\beta \omega \over 2} } \left|f_J\left(\Delta_H , \omega \right) \right|^2 |_{\eqref{eq:unitoperatorGFF}} = \theta(\omega) \sum_{n=0}^\infty c_{n,J} \delta(\omega - \Delta_L - 2n - J) , 
\ee
which is expected to become a good approximation at large spin $J \gg 1$. These are precisely the double-twist operators discussed in \cite{\citeHLLH}. The KMS image of the double-twist operators obtained by $\omega \to - \omega$, see \eqref{eq:ETHrelation} and \eqref{eq:KMSimage}, produces contributions regular in the $\tau \to 0$ limit (they are captured by the double-twist operators in the $t$-channel OPE).\footnote{For the same reason double-twist operators in the $t$-channel are sensitive to the boundary condition of the wave equation imposed at the horizon in the bulk.} 

It is clear, however, that in the present context such a conclusion would be too hasty. Indeed, let us imagine that instead of the discrete spectrum above we have a pair of closely separated poles that correspond to a pair of QNMs,
\be
\label{eq:doubletwistmirage}
&\delta \Big(\omega - \Delta_L - 2n - J - \gamma(n,J) \Big) \sim {1 \over 2 \pi i} \Big( {1 \over \omega - \Delta_L - 2n - J - \gamma(n,J) - i e^{- c_0(\mu) J }} - \text{c.c.} \Big),
\ee
where $\sim$ denotes equivalence in large $J$ perturbation theory.

A continuum with a pair of poles whose separation from the real axis is nonperturbatively small in spin is not distinguishable in large spin perturbation theory from the discrete sum. This is precisely what happens in our case! As we reviewed in Section \ref{radvstunn}, the QNMs acquire an imaginary part which is nonperturbative in spin due to the effect of tunneling. To conclude, the basic mechanism for reproducing the identity operator is different compared to \cite{Fitzpatrick:2012yx,Komargodski:2012ek} in the presence of the black hole horizon: in this case we have a continuum of operators (which are the black hole microstates), and ``double-twist operators'' are resonances whose imaginary part is nonperturbative in spin.

In fact, as we reviewed in Section \ref{radvstunn}, the nonperturbatively small imaginary part has the form $\exp(- c_0(\mu) J)$. The analysis in the previous works \cite{\citeHLLH} is organized perturbatively in $\mu \sim {\Delta_H \over c_T}$ and to leading order in $c_T$. It is then clear from \eqref{eq:doubletwistmirage} that in this perturbative expansion, the continuum disappears and the relevant spectral density becomes discrete. Therefore we can interpret the results of \cite{\citeHLLH} concerning the properties of double-twist operators $[\cO_H, \cO_L]$ as statements about the quasi-normal modes perturbatively in $\mu$, and correspondingly relate them to gravitational orbits studied in Section \ref{sec:classicalorbits}. We analyze this connection in detail in the next section.

In other words, if we  set ${\rm Im} \omega_{n,J}=0$ and $\lambda_{n,J} = - {i \over 2 \pi} e^{- {\beta \over 2} \omega_{n,J}} c_{n,J}$ in \eqref{eq:QNMexpansion}, we get the following representation of the correlator,
\be
\text{QNM}|_{{\rm Im} \omega_{n,J}, {\rm Re} \lambda_{n,J}=0} :~~~g(t, \theta) &=  \lambda_{H,H,L}^2 + 2\sum_{n,J}^\infty c_{n,J}  e^{ - \omega_{n,J} \beta/2 } \cos (\omega_{n,J} t) \ P_J\left( \cos \theta \right) ,
\ee
which is nothing but the standard OPE representation \eqref{eq:two-sided} where we sum over a discrete family of operators! This time, however, we effectively sum over CFT resonances instead of CFT operators. This is what happens as we work in the ${1 \over J} $ perturbation theory.

It is instructive to compare the ${1 \over J}$ expansion to the small $\mu$ expansion. In the latter case one can check that ${\rm Im} \omega_{n,J} \sim \mu^{J}$ \cite{cardoso,berti} and therefore the effects related to the double-twist operators emerging as the limit of resonances should appear at high enough order in $\mu$. It would be interesting to explore this aspect in detail.

\section{Double-twist dimensions from the Bohr-Sommerfeld formula}\label{sec:doubletwistdim}

For matter propagating on a black hole geometry, the concept of a normal mode is not applicable since waves can fall into the black hole. Instead one considers quasi-normal modes with complex energy $\omega$ \cite{Horowitz:1999jd,berti}. These are solutions to the wave equation which are purely ingoing at the horizon and satisfy normalizable boundary conditions at the boundary. By writing the solution to the wave equation in terms of the retarded Green function, one finds that the quasi-normal modes determine the late time behavior of the field. The real part of $\omega$ captures the oscillatory behavior of the wave, and the imaginary part of $\omega$ encodes the decay rate. \\ 
\indent Various tools have been developed for computing quasi-normal modes in different regimes. For our purposes, we are interested in quasi-normal modes that correspond to orbit states. This means that we take the mass of the orbiting particle to be large, so that the wavefunction is well-localized on the orbit. Since the spin $J$ of the state is proportional to $\Delta_L$, we are considering the large spin limit as well. This limit was considered in \cite{Festuccia:2008zx}, which we will review next. A similar large spin limit was analyzed for massless QNMs in the asymptotically flat case in \cite{Schutz:1985km}.

\subsection{Quasi-normal modes and the Bohr-Sommerfeld formula}
Given a scalar primary operator $\cO_L$ of dimension $\Delta_L$ we consider the dual scalar field $\phi$ in AdS of mass $m$. The two are related via the standard AdS/CFT dictionary \cite{Maldacena:2011ut}
\be
\Delta_L = {d \over 2} + \nu, ~~~ \nu = \sqrt{{d^2 \over 4} + m^2},
\ee
where we introduced a new parameter $\nu$. We will be interested in the semi-classical limit $\nu \gg 1$. 

The bulk field $\phi$ satisfies the Laplace equation on the black hole background. Using the symmetries of the problem, we can write 
\be
\phi(t,r, \vec n) = e^{- i \omega t} Y_{J,{\bf m}}(\vec n) \psi_{\omega, J}(r) ,
\ee
where $Y_{J,{\bf m}}(\vec n)$ are the spherical harmonics on $S^{d-1}$ of spin $J$.

\indent The quasi-normal modes are solutions to the radial wave equation with complex $\omega$ with the following boundary conditions 
\be
\psi_{\omega,J}(z) \simeq z^{{d \over 2} + \nu}, ~~~ z \to 0, \\
\psi_{\omega,J}(z) \sim e^{i \omega z}, ~~~ z \to \infty ,
\ee
where the latter condition means that the wave is purely ingoing at the horizon. Here the tortoise coordinate is defined by $dz=-dr/f(r)$ as in (\ref{tortoise}).
    
The wave equation for the radial part of the field $\psi_{\omega, J}(r)$ takes the form
\begin{align}
(-\partial_z^2+ \nu^2 V(z)-\omega^2)\psi_{\omega, J}(z)=0,\label{waveeq}
\end{align} 
where the potential $V(z)$ is
\begin{align}
V(z)=\frac{f(r)}{\nu^2}\left(\frac{(2J+d-2)^2-1}{4r^2}+\nu^2-\frac{1}{4}+\frac{\mu(d-1)^2}{4r^d}\right)\label{fullpotential},
\end{align}

We will be interested in solving the wave equation in the large $\nu$ limit or, equivalently, perturbatively in ${1 \over \nu}$ following \cite{Festuccia:2008zx}. Note the convenience of choosing $\nu$ as an expansion parameter (as opposed to $\Delta_L$) since the potential has a simple form $\#_0 + {\#_1 \over \nu^2}$. For the same reason it is convenient to choose a new parameterization for spin $J$
\be
2J+d-2=2\nu k .
\ee
Finally, we set
\begin{align}\label{scalinglimit}
\omega=\nu u , 
\end{align}
to get a nontrivial limit in the wave equation (\ref{waveeq}). The potential becomes
\begin{align}
V(z)=f(r)\left(\frac{k^2}{r^2}+1\right)+O(1/\nu^2).\label{potentiallimit}
\end{align}
Note that this matches the potential (\ref{eq:potentialBH}) describing the classical geodesic motion if we replace $k$ by $L$. As we will demonstrate shortly, this provides a direct link between the analysis of the present section and our discussion of the semi-classical orbits in Section \ref{sec:classicalorbits}.\\
\indent For large $\nu$, the wave equation (\ref{waveeq}) can be solved by a WKB analysis. The authors of \cite{Festuccia:2008zx} then showed how to compute the quasi-normal mode energies by analytically continuing the WKB wavefunctions in $u$. The resulting quasi-normal mode spectrum is discrete, and is quantized according to the Bohr-Sommerfeld rule. This rule becomes particularly simple when the potential has a minimum. This is the case when $k>k_{\text{min}}$, where $k_\text{min}$ is the critical momentum at which $V'(z)=V''(z)=0$ has a solution in $z$. This is what appeared as $J_{\text{min}}(\mu)$ in the analysis of the orbits in Section \ref{sec:classicalorbits}. For large black holes, we have
\begin{align}
k_\text{min}\sim \mu^{2/(d-2)},\hspace{10 mm}\mu\gg1.
\end{align}
\indent For $k<k_\text{min}$ the potential is monotonic, so the wave simply falls into the black hole, leading to an order one imaginary part of the quasi-normal mode energy. When $k>k_\text{min}$ there is a finite potential well, and the imaginary part of $\omega$ is related to the tunneling rate over the potential barrier, which is exponentially small, as we discussed in Section \ref{radvstunn}.  Neglecting these nonperturbatively small effects, for $k>k_\text{min}$ we have the usual Bohr-Sommerfeld rule for a particle moving in the potential,
\begin{align}
\nu \int_{z_a}^{z_b}dz\, \sqrt{u_n^2-V(z)}=\pi\left(n+1/2\right), \hspace{10 mm}n\ge 0. 
\label{leadingBS}
\end{align}
Here $z_{a}$ and $z_b$ are the turning points inside the potential well at which $u_n^2=V(z)$, see Figure \ref{potentialfig}. To leading order at large $\nu$ and ${n \over \nu}$ fixed \eqref{leadingBS} becomes the equation  \eqref{eq:BSclassical} from the classical orbit section. Formula \eqref{leadingBS}, however, is more precise in that it is applicable to finite $n$ as well. Next we discuss further $1/\nu$ corrections to \eqref{leadingBS}. \\
\indent

\subsection{Corrections to the Bohr-Sommerfeld formula}\label{sec:BScorrections}

The quantization condition (\ref{leadingBS}) allows us to compute the spectrum to leading order in $1/\nu$. However, it is possible to go further. Let us now give a systematic method for computing higher $1/\nu$ corrections. To do so we need to recall some little-known facts about higher order corrections to the Bohr-Sommerfeld rule in quantum mechanics \cite{PhysRev.41.713,Bender:1977dr}. \\
\indent We consider the second order differential equation
\begin{align}
\psi''(z)=\nu^2 Q(z) \psi(z),\hspace{10 mm}Q(z)\equiv V(z)-u^2\label{schrodinger}
\end{align}
where $\nu$ is a large parameter. We assume that $Q(z)$ has a unique minimum at $z=0$, and is monotonically increasing on both sides of this minimum. Then the spectrum of allowed values of $u$ is discrete. The wavefunction has a WKB expansion
\begin{align}
\psi(z)=\exp\left(\nu \sum_{i=0}^{\infty}\frac{1}{\nu^{i}}S_i\right).
\end{align}
Plugging into (\ref{schrodinger}), one finds that the $S_i$'s satisfy the recursion relations
\begin{align}\label{recursion}
S_0'(z)&=-\sqrt{Q(z)},\\
0&=2S_0'S_i'+\sum_{j=1}^{i-1}S_j'S_{i-1}'+S_{i-1}''.\
\end{align}
By matching the wavefunction near the turning points using an Airy function analysis, it is possible to derive the quantization relation to all orders in $1/\nu$ \cite{PhysRev.41.713}. This condition is 
\begin{align}
\frac{1}{2i}\oint dz\, \sum_{i=0}^{\infty}S_i'(z)=n\pi, \hspace{10 mm}n\ge 0,
\end{align}
where the contour integral is taken counterclockwise around the turning points. \\
\indent Let us explicitly write the quantization condition to first order in $1/\nu$. This is
\begin{align}
\nu\int_{z_a}^{z_b}dz\, \sqrt{-Q(z)}-\frac{1}{96 i\nu }\oint dz\, \frac{Q''(z)}{Q(z)^{3/2}}+O(1/\nu^2)=\pi(n+1/2).\label{BSfirstorder}
    \end{align}
Note that the replacement of $n$ with $n+1/2$ comes from the contour integral of $S_1'= - {1 \over 4} {d \log Q(z) \over d z}$, \cite{Bender:1977dr}
\be
{1 \over 2 i} \oint dz \Big( - {1 \over 4} {d \log Q(z) \over d z} \Big) = - {\pi \over 2},
\ee
where the counterclockwise integral of $\log Q(z)$ gives $2 \times 2 \pi i$ because it encircles a pair of simple zeros. 

Using the recursion relations (\ref{recursion}), corrections to any desired higher order in $\nu$ can be computed in terms of contour integrals of derivatives of $Q$.
\subsection{Perturbation theory around circular orbits}
The integrals on the left hand side of (\ref{BSfirstorder}) can be computed by perturbation theory around the minimum of the potential. Here we would like to consider a regime of $n$ in which this perturbation theory is valid, which allows us to obtain results to all orders in $\mu$ (see also \cite{Berenstein:2020vlp} for the leading order term in this expansion). Consider a nearly circular geodesic, which oscillates around the minimum of the potential with small amplitude. This means that the mode number $n$ is held fixed as $\nu\to \infty$. Then we can evaluate the integrals in perturbation theory around the minimum. We refer the reader to Appendix \ref{appendix} for the details. \\
\indent After performing the integrals, one finds the expression (\ref{finalquant}), which expresses the quantization condition for $n$ in terms of derivatives of the potential at the minimum. Now solving for $\omega=\nu u$ perturbatively in $\nu$, we have
\begin{align}
\omega &=\nu \sqrt{V}+\sqrt{\frac{V''}{2V}}(n+1/2)\label{BSfinal}\\
&\hspace{5 mm}+\frac{9(2n^2+2n+1)V''V''''V-(30n^2+30n+11)V'''^2V-18(2n+1)^2V''^3}{288\nu V^{3/2}V''^2}+O(1/\nu^2).\notag
\end{align}
In this expression, all derivatives are evaluated at the minimum of the potential. The power of this formula is that it is valid to \emph{all orders} in $\mu$, and also easy to evaluate explicitly. One simply needs to find the minimum of the potential and compute derivatives with respect to $z$. 

\subsection{Example: $O(\mu^3)$ in $d=4$}

\indent As an example of how to apply this technology, we consider the case $d=4$. The formulas for general dimensions are displayed in Appendix \ref{gammagend}. To match the known results from bootstrap, we solve for the minimum of the potential (\ref{fullpotential}) to second order in $\mu$ and $1/\nu^2$,
\begin{align}
r(z_{\text{min}})&=\left(1+\frac{k^2-1}{16k^2\nu^2}\right)\sqrt{k}-\frac{\mu}{k^{3/2}} \left(\frac{2k+1}{4}-\frac{2k^3+39k^2+70k}{64k^2\nu^2}\right)\\
&\hspace{10 mm}-\frac{\mu^2}{k^{7/2}}\left(\frac{28k^2+20k+3}{32}-\frac{84k^4+820k^3+1305k^2+620k-21}{512k^2\nu^2}\right)+O(\mu^3,1/\nu^3).\notag
\end{align}
\indent We now plug this value of the minimum radius into the expression for the energy $\ref{BSfinal}$. In doing so, one must be careful to use the full potential (\ref{fullpotential}) including $1/\nu^2$ corrections, and not just the leading order term. One finds
\begin{align}
\gamma_1&=-\frac{\nu^2+(6n+3)\nu+6n^2+6n+2}{2k\nu}+O(1/\nu^2),\label{gamma1} \\
\gamma_2&=-\frac{1}{8k^4\nu}\left( k(4k+1)\nu^2+k (21k+10)(2n+1)\nu\notag\right.\\
&\hspace{20 mm}\left.+35k^2+35k+4+ k (102k+90)n(n+1)\right)+O(1/\nu^2)\label{gamma2} \ .
\end{align}
Note that $n$ is held fixed as $\nu\to \infty$. As a special case of these formulas, we can take a circular orbit $n=0$ at leading order in $1/\nu$, 
\begin{align}
\gamma_1&=-\frac{\nu}{2k}+O(\nu^{0}),\hspace{10 mm}\gamma_2=-\frac{\nu (4k+1)}{8k^3}+O(\nu^0).
\end{align}
Comparing these results to \eqref{eq:gendgammaorbit} with $\alpha=0$ and $d=4$, one finds that the answers match at order $\nu$.\\
\indent Let us now compare with the results of the light-cone bootstrap. The value of $\gamma_1$ was computed at finite spin in  \cite{Li:2020dqm}, which found
\begin{align}
\gamma_1=-\frac{\Delta_L^2-\Delta_L+6n\Delta_L+6n^2-6n}{2(J+1)}.
\end{align}
Replacing $\Delta_L=2+\nu$ and $J=\nu k-1$, we see that this exactly matches the result (\ref{gamma1}). Note that the two results agree to all orders in $\nu$, even though our approximation is only valid to order $1/\nu$. One can check that this is a coincidence that only holds in $d=4$.\\
\indent At order $\mu^2$ and infinite spin, the anomalous dimension was computed in \cite{Kulaxizi:2018dxo,Li:2019zba},
\begin{align}
\gamma_2=-\frac{4\Delta_L^3+3(14n-1)\Delta_L^2+(102n^2-66n-1)\Delta_L+34(2n-1)n(n-1)}{8J^2}\label{gamma2him}.
\end{align}
Taking $k\to \infty$ in (\ref{gamma2}), we find that the answers again agree. Our formula $(\ref{gamma2})$ has additional finite spin corrections at order $O(1/J^3)$, which is a new prediction for the anomalous dimension $\gamma_2$ at finite spin. Note that the contribution of the final term in the numerator of (\ref{gamma2him}) is of order $1/\nu^2$, so we are not able to match it to this order in the Bohr-Sommerfeld approximation. \\
\indent To obtain another new prediction using the formula (\ref{BSfinal}), we can carry out the same process to order $\mu^3$. We omit the details since they are the same as above: one needs to solve for the minimum at order $\mu^3$, and compute derivatives of the potential to the same order. The result is 
\begin{align}
\gamma_3&=-\frac{1}{16k^5\nu} \Big( (4 k+1)^2 \nu ^2 +3 (41 k^2 + 34 k + 7 ) (2 n+1) \nu  \Big. \notag \\
&\Big. + 346 k^2 + 500 k + 199 + {16 \over k} + 12 ( 83 k^2+ 110 k + 35 ) n (n+1) \Big) . 
\end{align}
Note that it is relatively straightforward to generalize the expansion to higher orders in the small $\mu$ expansion. 

We present analogous formulas in $d$ spacetime dimensions in Appendix \ref{gammagend}.

\section{Gravitational orbits and many-body scars}
\label{sec:scars}

In this section we comment on the connection between gravitational orbits and the phenomenon of many-body scars recently discovered in the condensed matter systems \cite{bernien2017probing}, see also \cite{serbyn,moudgalya2021quantum} for reviews.\footnote{We thank Daniel Jafferis and Baur Mukhametzhanov for discussions on this topic.} Many-body scars are non-thermal energy eigenstates that violate ETH.
In this section, we demonstrate that the double-twist operators that appear in the light-cone bootstrap analysis of \cite{\citeHLLH} are scars, perturbatively in $1/J$.\footnote{In Appendix \ref{app:bulkhigherspin} we discuss the emergence of bulk higher spin symmetry at large spin, which naturally organizes the spectrum of the double-twist operators.} The presence of the black hole horizon (related to the nonperturbative in spin effects in the light-cone bootstrap) turns them into long-lived but eventually thermalizing states, similar to \cite{lin2020slow}. For this reason and due to the fact that their lifetime can be made arbitrarily big in the large $c_T$ limit, we can think of gravitational orbits as \emph{perturbative scars}.

\subsection{Five-point function}\label{fivepointfnc}

In this section we use the light-cone bootstrap technique to evaluate the one-point function of the light operator $\cO_L$ in the double-twist state that corresponds to a stable orbit. We imagine that the AdS theory contains a cubic coupling $g \phi^3$ which induces a nontrivial three-point function $\la \cO_L \cO_L \cO_L \ra \sim g$. To compute the one-point function we impose crossing on the five-point function that involves three light operators and two heavy operators.
We next consider a light-cone limit in which the two heavy operators become light-like separated. In this limit the leading contribution to the correlator comes from the factorized answer
\be
\label{eq:limit5}
\langle \cO_H(x_1) \cO_L(x_2)  \cO_L(x_3) \cO_L(x_5) \cO_H(x_4) \rangle & \overset{x_{14}^2 \to 0}{=} \langle \cO_H(x_1) \cO_H(x_4) \rangle \langle \cO_L(x_2)  \cO_L(x_3) \cO_L(x_5)  \rangle + ... \nn \\
&= {1 \over x_{14}^{2 \Delta_{H} } } {g \over (x_{23}^2 x_{25}^2 x_{35}^2)^{{\Delta_L/2}}} + ...  \ ,
\ee
where on the RHS we focused on the leading contribution as $x_{14}^2 \to 0$. We would like to understand how this result can be reproduced in the crossed channel $(12)$, $(34)$, where $(i j)$ stands for doing the OPE between $\cO_i (x_i) \cO_j (x_j)$. This setup is almost identical to the recent work \cite{Antunes:2021kmm} which we closely follow.

For simplicity we focus on the leading twist contribution only. To this end we write the following expansion in the light-cone limit,
\be
\label{eq:leadingtwist}
\cO_H(x_1) \cO_L(x_2)  \simeq \sum_{k} 2^J C_{H,L,k} \int_0^1 [d t]\, {\cO_{k,J}(x_1+t x_{21}, x_{12} ) \over (x_{12}^2)^{{\Delta_H + \Delta_L - \tau_k \over 2}}} , \notag\\
[d t] = {\Gamma(\Delta_k+J) t^{{\Delta_k +J - \Delta_H + \Delta_L \over 2} - 1} (1-t)^{{\Delta_k +J + \Delta_H - \Delta_L \over 2} - 1} \over \Gamma({\Delta_k+J \over 2} + {\Delta_H - \Delta_L \over 2})  \Gamma({\Delta_k+J \over 2} - {\Delta_H - \Delta_L \over 2})} ,
  \ee
where the RHS is designed to correctly reproduce the leading twist $\tau = \Delta - J$ part coming
from the operator $\cO_{k,J}(x)$ that appears in the OPE of $\cO_H \cO_L$.\footnote{We use the normalizations of the two- and three-point functions as in  \cite{Antunes:2021kmm}.}

Applying the formula \eqref{eq:leadingtwist} twice in the (12) and (45) channels we get 
\be
&\langle \cO_H(x_1) \cO_L(x_2)  \cO_L(x_3) \cO_L(x_5) \cO_H(x_4) \rangle \nn \\
&\simeq {1 \over (x_{12}^2 x_{34}^2)^{{\Delta_L+\Delta_H \over 2}}} \Big( {x_{13}^2 \over x_{15}^2 x_{35}^2 } \Big)^{{\Delta_L \over 2}}
\Big( {x_{23}^2 \over x_{14}^2} \Big)^{{\Delta_H- \Delta_L \over 2}}\sum_{k_1,k_2, \ell} P_{k_1 k_2 \ell} {\cal G}_{k_1 k_2 \ell}(u_i) , 
\ee
where $k_i$ label operators that appear in the heavy-light channels \eqref{eq:leadingtwist}, and the quantum number $0 \leq \ell \leq \min ( J_{k_1}, J_{k_2})$ labels different tensor structures in the three-point function $\langle \cO_{k_1} \cO_L \cO_{k_2} \rangle$.

We have also introduced $P_{k_1 k_2 \ell}$, defined as a product of the three-point functions
\be
P_{k_1 k_2 \ell} = C_{H,L,k_1} C_{H,L,k_2} C_{k_1, k_2, \ell} ,
\ee
and $C_{k_1, k_2, \ell}$ describes the three-point function of the light operator and two orbit states $\langle \cO_{k_1} \cO_L \cO_{k_2} \rangle$ that we want to compute. 

The collinear blocks ${\cal G}_{k_1 k_2 \ell}(u_i)$ take the form
\be
 &{\cal G}_{k_1 k_2 \ell}(u_i)=u_1^{{\tau_1 \over 2}} u_3^{{\tau_2 \over 2}} (1-u_2)^{\ell} u_5^{{\Delta_L \over 2}} u_2^{{\Delta_L - \Delta_{H} \over 2}} \int_0^1 [d t_1] [d t_2] \nn \\
 &\times {\left( 1 + t_1 (1-u_2) u_4 - u_4 \right)^{J_2 - \ell} \left( 1 + t_2 (1-u_2) u_5 - u_5 \right)^{J_1 - \ell}  \over (u_4+ t_2 (1-u_4) )^{{\Delta_2-\Delta_1+ J_1 + J_2 -2 \ell + \Delta_L \over 2}} (u_5 + t_1 (1-u_5) )^{{\Delta_1-\Delta_2+ J_1 + J_2 -2 \ell + \Delta_L \over 2}} } \nn \\
 &\times {1 \over (1- t_1 t_2 (1-u_2))^{{\Delta_1+\Delta_2 + J_1 + J_2 - \Delta_L \over 2}}} ,
\ee
where $0 \leq \ell \leq \min (J_1,J_2)$ labels various three-point function tensor structures in the spinning correlator $\la \cO_{J_1} \cO_{L} \cO_{J_2} \ra$, see \cite{Antunes:2021kmm} for details. In the formula above we introduced conformal cross-ratios as follows,
\be
u_i = {x_{i,i+1}^2 x_{i+2,i+4}^2 \over x_{i,i+2}^2 x_{i+1,i+4}^2}~, ~~~ i \in \mathbb{N} ~(\text{mod} ~5) .
\ee
Setting $\Delta_H = \Delta_L$ the result above reproduces the corresponding formula from \cite{Antunes:2021kmm} (upon doing the change of variables $t_i \to 1-t_i$).

To reproduce \eqref{eq:limit5} we thus get the following equation,
\be
\label{eq:crossing5}
\sum_{k_1,k_2,J} P_{k_1 k_2 J} {\cal G}_{k_1 k_2 J}(u_i) =g (u_1 u_3)^{{\Delta_H+\Delta_L \over 2}} u_2^{-{\Delta_H \over 2}} u_5^{{\Delta_L \over 2}}. 
\ee
Matching the dependence on $u_1$ and $u_3$ we get $\tau_1 = \tau_2 = \Delta_H+\Delta_L $ which is simply the statement that the result is reproduced by the leading twist double-twist operators. 

At this point we can take the limit $\Delta_H \to \infty$ and also introduce $J_i = \ell + j_i$. Under the integral we set $\Delta_i = \Delta_H+\Delta_L+J_i$ and rescale $t_i \to {t_i \over \Delta_H}$, after which both the limit and the integral can be trivially computed. The crossing equation \eqref{eq:crossing5} then takes the form
\be
\sum_{\ell=0}^\infty (1-u_2)^{\ell} \sum_{j_1,j_2=0}^\infty (u_4^{-1}-1)^{j_2} (u_5^{-1}-1)^{j_1} P_{\ell, j_1, j_2} =g \left( {u_4 u_5 \over u_2} \right)^{{\Delta_L \over 2}} + ...  , ~ u_2 \to 0 . 
\ee
Note that the dependence on $u_2$ is non-analytic around $u_2=0$ on the RHS and analytic for fixed $\ell$ on the LHS. This means that it can only be generated by the infinite spin tail.

Let us introduce
\be
g_{\ell}(u_4, u_5) = \sum_{j_1,j_2=0}^\infty (u_4^{-1}-1)^{j_2} (u_5^{-1}-1)^{j_1} P_{\ell, j_1, j_2} . 
\ee 
A simple way to satisfy crossing is to impose the following condition,
\be
\lim_{\ell \to \infty} g_{\ell}(u_4, u_5) = g \left( u_4 u_5  \right)^{{\Delta_L \over 2}}{\ell^{{\Delta_L \over 2} - 1} \over \Gamma({\Delta_L \over 2})} . 
\ee
In this way we get
\be
\lim_{\ell \to \infty} P_{\ell, j_1, j_2} = (-1)^{j_1 + j_2} g { \ell^{{\Delta_L \over 2} - 1} \over \Gamma({\Delta_L \over 2})} {\Gamma(j_1+{\Delta_L \over 2}) \over \Gamma(j_1+1) \Gamma({\Delta_L \over 2})}{\Gamma(j_2+{\Delta_L \over 2}) \over \Gamma(j_2+1) \Gamma({\Delta_L \over 2})}.
\ee
Finally, dividing by the GFF three-point functions of the double-twist operators we get for the desired one-point function\footnote{Here we have chosen to define the heavy-light double twist operators of odd spin in a way that the corresponding GFF three-point function contains $(-1)^J$.}
\be
C_{j_1, j_2, \ell} = {g \over \ell^{{\Delta_L \over 2}} } {\Gamma(\Delta_L) \over \Gamma({\Delta_L \over 2})^3} {\Gamma(j_1+{\Delta_L \over 2}) \Gamma(j_2+{\Delta_L \over 2}) \over \Gamma(j_1+1)  \Gamma(j_2+1) } , ~~~ J_i = j_i + \ell, ~~~ \ell \gg 1 . 
\ee

As expected, the one-point function of the light operator in the double-twist state is different from zero to leading order in $c_T$, given a non-zero three-point function $\la \cO_L \cO_L \cO_L \ra \sim g$. In contrast it is zero in the state which is dual to the rotating Kerr black hole with the same quantum numbers. Perturbatively in spin these two states do not mix, and thus orbit states represent eigenstates that naively violate ETH. However, the nonperturbative effect of tunneling introduces mixing between the orbit states and the black hole states. Orbits become long-lived resonances which eventually thermalize.

\subsection{Perturbative scars}\label{perturbativescars}

\indent In any chaotic theory, the matrix elements of simple operators in energy eigenstates are expected to take the ETH form (\ref{eq:ETH}). In some systems the ETH is known to hold in all energy eigenstates \cite{kim2014testing}. However, recently a class of systems has been discovered, where the ETH holds in all but a small number of eigenstates \cite{bernien2017probing,turner}. Eigenstates in the middle of the spectrum whose matrix elements violate the ETH are called quantum scars. Typically they occupy a small subsector of the Hilbert space but lead to interesting phenomena such as revivals and lack of thermalization. Many examples of theories containing scars have been found, and we refer the reader to the reviews \cite{moudgalya2021quantum,serbyn} for a survey of the literature on the subject. Many of the models that have been found can be understood based on symmetry properties of the Hamiltonian \cite{Pakrouski:2020hym,Pakrouski:2021jon}. 

\indent Although there are theories containing isolated scar states, some of the simplest cases contain towers of scars with equally spaced energy eigenstates. For example, the Hubbard model contains states with equally spaced energy of the form $(\eta^\dagger)^n|0\rangle$ \cite{PhysRevLett.63.2144}. Here $\eta^\dagger$ is the raising operator for a pseudospin $SU(2)$ symmetry, and raises the energy of the state by a fixed amount. It is possible to add an interaction that breaks the symmetry but preserves the tower of states \cite{PhysRevB.102.075132,PhysRevB.102.085140}, so that these states constitute a tower of many-body scars. In this example the pseudospin $SU(2)$ is referred to as a spectrum generating algebra, since the full tower can be generated by acting with $\eta^\dagger$ on the vacuum. 

\indent Now let us turn to the case of holographic CFTs. In this paper we discussed the following interesting phenomenon. Starting from the four-point function of heavy-light operators and performing the light-cone bootstrap analysis \cite{\citeHLLH}, one concludes that the spectrum contains an infinite family of double-twist states. The light-cone bootstrap analysis of the five-point function (along the lines of \cite{Antunes:2021kmm}) reveals that these double-twist states violate ETH while being in the middle of the spectrum \eqref{eq:regionofM}, and as such they look like scars. In the bulk these non-thermal states correspond to gravitational orbits around black holes.

The bulk picture, however, immediately reveals the limitation of this conclusion: the gravitational orbits are not stable due to tunneling and gravitational radiation. Therefore gravitational orbits are long-lived states that eventually thermalize. Both effects are closely related to the presence of the black hole horizon in the bulk. Its presence also indicates that the spectrum of the dual CFT is effectively continuous. How is it possible then that the light-cone bootstrap analysis reveals a set of discrete states?

The resolution of this puzzle is that the continuum of states contains an infinite set of narrow resonances whose widths is nonperturbative in ${1 \over J}$, more precisely $\exp(- c_0(\mu) J)$, see formulas (\ref{tunnsmallmu}) and (\ref{tunnlargemu}). The light-cone bootstrap large spin expansion studied in \cite{\citeHLLH} misses such effects, and as we explained the set of resonances (or quasi-normal modes) becomes the set of double-twist operators that we described above. This can also be understood by recalling that at $J=\infty$ the spectrum of the CFT is effectively controlled by the bulk higher spin symmetry, and the double-twist operators form a multiplet under this symmetry, see Appendix \ref{app:bulkhigherspin}. As we go to finite spin $J$ the bulk higher spin symmetry gets broken by $\epsilon = {1 \over J}$ effects. Perturbatively in $\epsilon$, the double-twist operators persist as non-thermal energy eigenstates. Nonperturbatively in $\epsilon$ they disappear from the spectrum and become long-lived resonances.

In this sense we can say that double-twist operators (or gravitational orbits) present an example of \emph{perturbative scars:} an infinite family of long-lived states whose lifetime can be made arbitrarily big in the large $c_T$ limit and whose lifetime is nonperturbative in the symmetry breaking parameter $\epsilon$. \\
\indent One notable difference between the states we have found and conventional examples of quantum scars is that the latter have a sub-volume law scaling of the entanglement entropy. The towers of scars analyzed in the literature generally consist of quasiparticles above a low-entanglement state, which implies that the states in the tower have a smaller entanglement than a state obeying ETH. However, there are also known examples with a volume-law scaling \cite{Langlett:2021efq}. In our case, the presence of the black hole means that the entanglement entropy scales with the volume. \\
\indent The discussion so far applies to perturbation theory in $1/J$. At finite $J$, the orbits are no longer exact eigenstates, but are instead broadened into resonances. They can be expressed as a sum of exact eigenstates in a band of width $\Gamma$,
\begin{align}
|\text{orbit}\rangle=\sum_{E_i=E_{\text{orbit}}-\Gamma}^{E_{\text{orbit}}+\Gamma}c_i|E_i\rangle,\hspace{10 mm}\sum_{i=1}^N |c_i|^2=1.
\end{align}
Here we have defined $N\sim e^{S(E_{\text{orbit}})}$, and the normalization condition implies that $|c_j| \sim O(N^{-1/2})$. Let us assume that each of the eigenstates $|E_i\rangle$ obeys ETH. Then the one-point function of a light operator in the orbit state is
\begin{align}
\langle \text{orbit}|\cO_L|\text{orbit}\rangle&= \sum_{i,j} c_j^* c_i \langle E_j | \cO_L|E_i\rangle\notag \\
&= \sum_{i,j} e^{-S \Big({E_i + E_j \over 2} \Big)/2} c_j^* R_{j,i} c_i  f_{\cO_L} \Big({E_i + E_j \over 2}, E_i - E_j\Big) ,
\end{align}
where we used that by assumption the one-point function of $\cO_L$ in each energy eigenstates is suppressed at large $c_T$. Let us write the pseudorandom matrix $R_{j,i} =|R_{j,i}| e^{i \phi_{j,i}}$, where hermiticity implies that $\phi_{j,i} = - \phi_{i,j}$ and $|R_{j,i}| \sim O(1)$. We can also write $c_i = |c_i| e^{i \phi_i}$.  We would like to make the matrix element $O(1)$ by choosing the   phases in a way that they add up instead of canceling each other. Note that there are $O(N^2)$ phases $\phi_{i.j}$, so we cannot cancel all of them since we only have $O(N)$ coefficients at our disposal. 

Let us consider a simplified model where we set $f_{\cO_L}(\bar E, \omega)=1$ in the model above. We also consider a state where $|c_i|=\frac{1}{\sqrt{N}}$ for all $i$. We first estimate the sum over $j$,
\be
\sum_{j=1}^{N} R_{j,i} c_j^* \equiv O(1) \times e^{- i \tilde \phi_i} ,
\ee
where the $O(1)$ coefficient comes as follows: the sum over $N$ random phases produces $\sqrt{N}$, which together with the fact that $|c_j| \sim O(N^{-1/2})$ gives something $O(1)$. 

We next choose the phases of $c_i$ to be
\be
\phi_i = \tilde \phi_i, ~~~ i = 1, ..., N .
\ee
This produces the following estimate,
\be
\sum_{i,j} e^{-S \Big({E_i + E_j \over 2} \Big)/2} c_j^* R_{j,i} c_i &\simeq e^{- S(E_{\text{orbit}})/2} \sum_{i,j=1}^N c_j^* R_{j,i} c_i \nn \\
&\sim e^{- S(E_{\text{orbit}})/2} \sum_{i=1}^N c_i e^{- i \tilde \phi_i} \notag\\
&\sim  O(N^{1/2}) e^{- S(E_{\text{orbit}})/2}\sim O(1),
\ee
where we used $N \sim e^{S(E_{\text{orbit}})}$.  Therefore we conclude that the matrix elements $O(1)$ are indeed consistent with the ETH upon a proper choice of $c_i$. In this way, at finite $c_T$ and $J$ we can think of orbits as superpositions of the black hole microstates.

\indent It is instructive to contrast the orbit states with other long-lived quasi-normal modes. For instance, the large AdS black hole has a family of parity-odd gravitational quasi-normal modes with purely imaginary frequency, \cite{cardoso,berti}
\begin{align}\label{longlivedj}
\omega_J=-\frac{i}{r_s}\frac{(J-1)(J+d-1)}{d}, ~~~ r_s \gg 1.
\end{align}
The decay time of these modes is proportional to $r_s$, which is much longer than the expected thermalization time $1/r_s$ at high temperatures. Therefore these modes are long-lived resonances, just like the orbit states. However, there are several key differences between the modes (\ref{longlivedj}) and the orbits: 
\begin{itemize}
    \item Since $\text{Re }\omega=0$ in (\ref{longlivedj}), the overlap $\langle \omega_J(t)|\omega_J(0)\rangle$ decays exponentially in time but does not oscillate. This is in contrast to the orbit modes, for which $\omega$ has a nonzero real part, and which exhibit approximate revivals.
    \item In the case of the orbits the decay rate decreases exponentially with spin, while the decay rate of the modes (\ref{longlivedj}) grows polynomially in spin. In particular, for (\ref{longlivedj}) there is no small parameter analogous to $\mu/J^{d/2-1}$ in which to expand. Therefore there is not a limit in which these states become approximate energy eigenstates for which we can identify the corresponding boundary operator.
    \item The orbit quasi-normal modes come in a two-parameter family labeled by $J$ and $n$, while the modes (\ref{longlivedj}) are only parameterized by $J$. In contrast to the orbit states, they also disappear from the spectrum in the $\mu \to 0$ limit \cite{cardoso,berti}.
\end{itemize}

\section{Discussion}

In this paper we have considered classically stable orbits around AdS Schwarzschild black holes. These are long-lived states in the bulk that eventually decay due to tunneling and gravitational radiation. The tunneling rate is given by $e^{-c_0(\mu)J}$, where $J$ is the spin of the orbit. We have computed the life-time of the orbits due to scalar radiation in the case of large black holes, $\mu \gg 1$ and large spin $J \gg \mu^2$, and found that it is $\sim {J^4 \over  {\kappa^2\mu^8}}$ in $d=3$. In the analogous gravitational problem, the lifetime scales as $\sim c_T$.

The existence of such orbits in the bulk have various manifestations in the dual conformal field theory \cite{Festuccia:2008zx,Berenstein:2020vlp}. First, they correspond to quasi-normal modes in the thermal two-point function \cite{Festuccia:2008zx}
\be
\text{Gravitational orbits} ~~~\leftrightarrow~~~ \text{Quasi-normal modes} \ . \nn 
\ee
The fact that they are long-lived is mapped to the fact that the imaginary part of the corresponding quasi-normal modes is very small. Second, via the ETH they appear as resonances in the heavy-light OPE, where the heavy operator is dual to the black hole, and the light operator is dual to the orbiting body. By resonances we mean poles on the second sheet of the conformal partial waves $c(\Delta,J)$.
It is then natural to consider the expansion of the heavy-light four-point function in terms of the QNMs \eqref{eq:QNMexpansion}.

After connecting orbits to QNMs and considering the QNM expansion of the heavy-light four-point correlator we noted that there is a natural expansion in which QNMs \emph{look} like energy eigenstates or primary operators. First, we take the large $c_T$ limit. Second, we consider the large $J$ expansion. In this setting, orbits become stable: tunneling is nonperturbative in $J$; gravitational radiation is ${1 \over c_T}$ suppressed. In fact this is precisely the setup studied in the light-cone bootstrap \cite{Fitzpatrick:2012yx,Komargodski:2012ek} that has been recently applied to the heavy-light correlators \cite{\citeHLLH}. The orbital quasi-normal modes then become nothing but the double-twist operators\footnote{The relation between the double-twist operators and stable orbits has previously been discussed in \cite{Berenstein:2020vlp}.}
\be
\text{Quasi-normal modes} ~~~\leftrightarrow~~~ \text{Double-twist operators} \ . \nn
\ee
We then used the Bohr-Sommerfeld quantization formula and corrections thereto to compute the anomalous dimensions of the double-twist operators in the semi-classical expansion: ${1 \over \Delta_L} \to 0$, ${J \over \Delta_L}$ fixed. We found complete agreement with the results obtained using the light-cone bootstrap and we made further all-order in $\mu$ predictions to the anomalous dimensions of the double-twist operators, see e.g. Figure \ref{fig:ReggeTr}.

Finally, following \cite{Antunes:2021kmm}, we used the five-point light-cone bootstrap analysis of the heavy-light operator to compute the one-point function of a light operator in the double-twist states. As expected based on the correspondence of the double-twist states with orbits in AdS, the one-point functions are not thermal. Such a behavior was recently observed in many condensed matter systems and the corresponding phenomenon is known as many-body scars \cite{serbyn,moudgalya2021quantum}. We thus find that there is a natural connection
\be
\text{Double-twist operators}~~~\leftrightarrow~~~ \text{Many-body scars} \ . \nn 
\ee
However, there is an important difference between the phenomenon of scars observed in condensed matter systems and in holography. In our case the non-thermal nature of the orbits is an artifact of working in perturbation theory in $1/J$ and to leading order in $c_T$. Including nonperturbative effects in $1/J$ or $1/c_T$ corrections turn double-twist operators into resonances, and make the spectrum continuous as opposed to being discrete. In this sense we can call double-twist operators \emph{perturbative scars}. \\
 \indent The correspondence between orbits, QNMs, double-twist operators, and scars can be succinctly summarized by the equation
 \begin{align}
 \label{eq:finalformula}
 {1 \over 2 \pi i} \Big(\overbrace{ {1 \over \underbrace{\Delta - \Delta_n(J)}_{\text{stable orbit}} - i e^{- c_0(\mu) J}} - {1 \over \Delta - \Delta_n(J) + i \underbrace{ e^{- c_0(\mu) J}}_{\text{BH tunneling}} } }^{\text{BH microstates}}\Big) \overset{\text{PT}}{\simeq} \overbrace{\underbrace{\delta(\Delta - \Delta_n(J) )}_{\text{double-twist}}}^{\text{scar}},
 \end{align}
 where PT denotes that the equivalence holds in perturbation theory in $1/J$. Of course, the continuum on the LHS of \eqref{eq:finalformula} is a large $c_T$ limit effect. In particular, we always work in the regime $e^{- c_0(\mu)J} \gg e^{- c_T}$, where $e^{- c_T}$ is the scale associated with the discreteness of the CFT spectrum.

Let us comment on a few possible future directions: 

\begin{itemize}
\item In this paper we have focused on the simplest case of Schwarzschild-AdS black holes, but orbits should exist in more general situations as well. It would be interesting to generalize our analysis to other gravitational solutions, such as charged, rotating, extremal, and supersymmetric black holes. On the boundary, this corresponds to considering $[\mathcal{O}_H\mathcal{O}_L]_{n,J}$, where now $\mathcal{O}_H$ is a more general heavy operator with one of the aforementioned properties. It would be interesting to use the Bohr-Sommerfeld condition to compute the spectrum of anomalous dimensions in this more general case, as well as to reproduce these results using the light-cone bootstrap.
\item Quasi-normal modes of black holes in flat space were recently connected to four-dimensional supersymmetric gauge theories \cite{Aminov:2020yma}, see also \cite{Bianchi:2021mft,Bonelli:2021uvf}. In that work an exact Bohr-Sommerfeld quantization condition was formulated and solved using the Nekrasov partition function in a particular phase of the $\Omega$-background \cite{Nekrasov:2002qd,Nekrasov:2003rj,Nekrasov:2009rc}. This connection can be also generalized to AdS black holes \cite{AlbaWIP}. It would be very interesting to explore this connection further in the context of the conformal bootstrap and see if it can be used to ``solve'' the thermal two-point function in the black hole background. 
\item It would be interesting to develop a deeper understanding of the connection between the many-body scars and gravitational orbits. In particular, it would be very interesting to see if there are other condensed matter systems which exhibit a similar phenomenon of perturbative scars.\footnote{PXP scars \cite{turner,PhysRevLett.122.040603,PhysRevX.11.021021} are believed to eventually thermalize at very long time scales. If this were true, then the PXP model would provide another example of approximate quantum scars. We thank Zlatan Papi\'{c} for explaining this to us.} As emphasized in \cite{Berenstein:2020vlp}, the spatial curvature and the finite volume of the space on which the quantum system lives are necessary for the existence of stable orbits in the gravity dual. It would be also very interesting to understand the interplay between the lifetime of gravitational orbits and maximal chaos. In the context of holographic theories this is related to understanding the fate of gravitational orbits at finite 't Hooft coupling $\lambda$ and understanding how stringy corrections affect the lifetime of the orbits.
\item We have observed that a finite lifetime of the gravitational orbits drastically changes the structure of the heavy-light OPE. Instead of a discrete sum over the double-twist operators we get a continuum spectrum with many narrow resonances. The width of these resonances is nonperturbative in spin $J$. It would be interesting to explore nonperturbative in spin $J$ effects using the Lorentzian inversion formula, which should correctly capture them \cite{Caron-Huot:2017vep,Caron-Huot:2020adz}. It would be also interesting to see if the heavy-light four-point function bootstrap together with the ETH could provide new insights into the finite temperature bootstrap \cite{Iliesiu:2018fao,Iliesiu:2018zlz,Alday:2020eua}.
\item Existence of gravitational orbits around AdS black holes is a very robust feature of holographic theories. In particular, it would be interesting to analyze orbits when the geometry of the boundary is different from $S^{d-1}$. It is clear that the positive curvature of $S^{d-1}$ is important for having gravitational orbits, e.g. they are obviously absent for $\mathbb{R}^{d-1}$ or $T^{d-1}$. Relatedly, it would be interesting to better understand the implication of the light-cone bootstrap for CFTs on general spatial manifolds $\mathbb{R} \times \Sigma$.
\item In this work we have focused on gravitational orbits in asymptotically AdS spaces. As Earthlings well know,\footnote{This comment does not apply to flat-Earthers.} stable orbits are characteristic to the gravitational dynamics in four dimensions in asymptotically flat and de Sitter spacetimes as well. It would be very interesting to understand how potentially very intricate structure of the orbits, e.g. the Milky way galaxy, is realized in the dual theories and if there are simple toy models that could correctly capture the gross features of the orbital dynamics (together with the maximal chaos).

\item An interesting aspect of the heavy-light bootstrap is the role of the horizon in the dual classical geometry. The presence of the horizon makes the spectrum of the normalizable solutions to the bulk wave equation continuous (and correspondingly the spectrum of the dual CFT). Instead having a horizon-less geometry, e.g. an AdS star \cite{deBoer:2009wk,Arsiwalla:2010bt}, would produce a discrete spectrum in the heavy-light channel. Both geometries look identical close to the boundary (due to the no-hair theorem) and in a related manner they will acquire an identical contribution from the multi-trace stress energy tensor operators $T^k$ as discussed, for example, in \cite{Fitzpatrick:2019zqz}. The difference between the two geometries is captured in the light-light channel by the properties of the double-twist operators $\cO_L \Box^n \partial^J \cO_L$ as well as by the spectrum of the double-twist heavy-light operators $\cO_H \Box^n \partial^J \cO_L$. When using the Lorentzian inversion formula \cite{Caron-Huot:2017vep,Li:2019zba,Li:2020dqm} the contribution of the double-twist operators $\cO_L \Box^n \partial^J \cO_L$ is suppressed by ${1 \over c_T}$, but the contribution of the operators $\cO_H \Box^n \partial^J \cO_L$ is only suppressed by powers of $\mu$. Therefore, the difference between black holes and stars will be visible (see \cite{Giusto:2020mup,Ceplak:2021wak} for related work in the $AdS_3$ context). It would be interesting to explore these effects in detail.
\item It would be very interesting to generalize our discussion to finite $c_T$. There are many places in which our discussion will have to be modified. One important effect is gravitational radiation, which contributes to the lifetime of the orbits at order $1/c_T$. More conceptually, the basic features of the black hole geometry, such as the black hole horizon or the black hole singularity \cite{Festuccia:2005pi,Festuccia:2006sa,Festuccia:2008zx}, naturally appear on the second sheet of the conformal partial wave expansion $c(\Delta,J)$. The notion of a second sheet of $c(\Delta,J)$ is a large $c_T$ effect, which is absent in a single CFT at finite $c_T$ with a  discrete spectrum. Still, it should be possible to define the second sheet of $c(\Delta, J)$ at finite $c_T$ upon a proper coarse-graining procedure. Naturally it should be related to the experience of a low-energy observer in the bulk with a finite energy resolution. For example, it is natural to smear $c(\Delta,J)$ over a finite region of the $\Delta$-plane, which effectively creates a cut even at finite $c_T$, see e.g. \cite{Mukhametzhanov:2018zja}. Indeed, perturbative in $c_T$ computations effectively perform such an averaging in a region of size $1/c_T^{\#}$ since they do not resolve the $e^{-c_T}$ discreteness of the spectrum. It would be also interesting to explore the effects of other notions of averaging in higher-dimensional CFTs that have been recently discussed in the literature, see e.g. \cite{Pollack:2020gfa,Belin:2020hea,Collier:2022emf,Schlenker:2022dyo,Chandra:2022bqq,Heckman:2021vzx}.
\end{itemize}

\section*{Acknowledgements}
We thank Alexandre Belin, Shouvik Datta, Anatoly Dymarsky, Thomas Iadecola, Alba Grassi, Daniel Jafferis, Daniel Kapec, Shota Komatsu, Baur Mukhametzhanov, Kyriakos Papadodimas, G\'abor S\'arosi,  Wilke van der Schee, Steven Shenker, and Evgeny Skvortsov for helpful discussions. We thank Liam Fitzpatrick for comments on the manuscript. This project has received funding from the European Research Council (ERC) under the European Union’s Horizon 2020 research and innovation programme (grant agreement number 949077). 

\appendix
\section{Radiation from non-circular orbits}\label{radnoncircular}
In this appendix we will consider radiation from a general stable orbit, which oscillates in the radial direction.  This generalizes the classic calculation of \cite{Peters:1963ux,Peters:1964zz}, which considered gravitational radiation from Keplerian orbits with nonzero eccentricity. We consider the case where the turning points are far away from the ISCO, $r_{a},r_b\gg \mu$. In this limit the geodesic can be approximated by an AdS geodesic, which is an ellipse with eccentricity $x=\sqrt{1-(r_b/r_a)^2}$,
\begin{align}
r(t)=\frac{r_b}{\sqrt{1-x^2\sin^2t}}.\label{ellipse}
\end{align}
Proceeding as in Section \ref{radcircular}, we find that the solution near the horizon is
\begin{align}
\psi_{Jm\omega}(z)&\sim\frac{\kappa e^{-i\omega z}}{2i \omega C_{J\omega}}Y_{J m}^*(\pi/2, 0)\int_{-\infty}^{\infty} dt\, \frac{e^{i(\omega t-m\phi(t))}}{r(t)^3},\hspace{10 mm}z\to \infty.\label{noncircularintegral}
\end{align}
\indent We now plug the elliptical geodesic motion (\ref{ellipse}) into the integral in (\ref{noncircularintegral}),
\begin{align}
\int_{-\infty}^{\infty} dt\, \frac{e^{i(\omega t-m\phi(t))}}{r(t)^3}=\frac{1}{r_-^3}\int_{-\infty}^{\infty} dt\,e^{i(\omega -m)t}(1-x^2\sin^2 t)^{3/2}.
\end{align}
This is the Fourier transform of a periodic function with period $\pi$, so it takes the form 
\begin{align}
\frac{1}{r_-^3}\sum_{n=-\infty}^{\infty}c_{ n}\delta(\omega-m-2n),\hspace{10 mm}n\in \mathbb{Z}.
\end{align}
The Fourier coefficients are
\begin{align}
c_{-n}=c_n&=\frac{1}{\pi }\int_{0}^{\pi}dt\, e^{2int}(1-x^2\sin^2t)^{3/2}\\
&=\frac{3(-1)^n}{4^{n+1}\sqrt{\pi}}\frac{\Gamma(n-3/2)}{\Gamma(n+1)}x^{2n}{_2F_1}(-3/2+n,1/2+n,1+2n,x^2),\hspace{10 mm}n\ge 0\notag.
\end{align}
\indent There are two different interesting limits. The first regime is an almost circular orbit. Then $r_+\sim r_-$, so we find 
\begin{align}
&\int_{-\infty}^{\infty} dt\,e^{i(\omega -m)t}\left(1+\frac{3}{2}x\sin^2 t\right)=\delta(\omega-m)-\frac{3x}{8}(\delta(\omega-m+2)+\delta(\omega-m-2)).\label{cnsmallx}
\end{align}
This means that we have started exciting energies close to $\omega=m$. The second regime is a highly eccentric orbit. Then the Fourier coefficients are 
\begin{align}
c_{n}&=\frac{12(-1)^{n}}{\pi (9-40n^2+16n^4)}.
\end{align}
\indent Now let us examine the power output. This is time-dependent, but we can average over one period to compute the average energy and angular momentum loss, 
\begin{align}
\left\langle \frac{dE}{dt}\right\rangle&=-\sum_{J m n}\frac{\kappa^2|Y_{J m}(\pi/2,0)|^2|c_{n}|^2}{|C_{J,m+2n}|^2r_-^6}\\
\left\langle \frac{dL}{dt}\right\rangle&=-\sum_{J m n}\frac{\kappa^2|Y_{J m}(\pi/2,0)|^2|c_{n}|^2}{|C_{J,m+2n}|^2(1+2n/m)r_-^6}.
\end{align} 
We will not attempt to obtain a general solution of these differential equations, but let us give an approximate solution for small eccentricity. Since the sum is dominated by small $n\ll m$, we have from (\ref{cnsmallx})
\begin{align}
\left\langle \frac{d(E-L)}{dt}\right\rangle&\sim -\sum_{J m n}\frac{n^2}{m^2}\frac{\kappa^2|Y_{lm}(\pi/2,0)|^2|c_{n}|^2}{|C_{J,m+2n}|^2r_-^6}\notag\\
&\sim -\frac{x^2 \kappa^2 \mu^{4}}{r_-^6}\label{eminusldot}.
\end{align}
On the other hand, for small $x$ the orbit equations (\ref{eq:noncircQN}) give
\begin{align}
E-L=1+\frac{x^4}{8}.\label{eminusl}
\end{align}
Comparing (\ref{eminusl}) with (\ref{eminusldot}), we find
\begin{align}
\dot{x}\sim -\frac{\kappa^2\mu^4}{xr_-^6}.
\end{align}
For very small $x$, we can neglect the variation in $r_-$ over time. Therefore we find that the orbit becomes circular in a time 
\begin{align}
t_{\text{circular}}=\frac{r_-^6x_0^2}{\kappa^2 \mu^4}\ll t_{\text{decay}}.
\end{align} 
\section{Corrections to Bohr-Sommerfeld quantization}\label{appendix}
\begin{figure}[t]
\centering
\includegraphics[scale=.45]{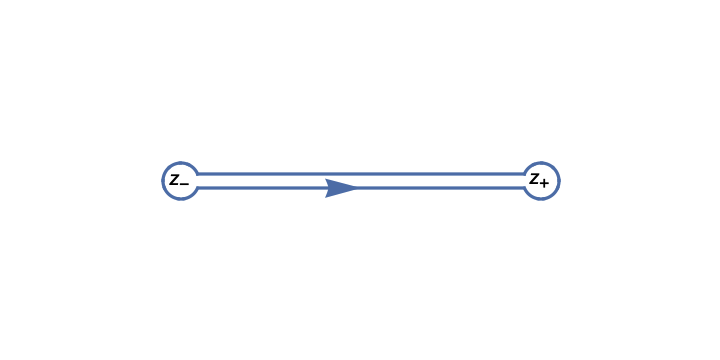}
\caption{We choose a counterclockwise contour that is straight except near the turning points $z_{\pm}$, where the contour is an infinitesimal circular arc. The integral over the straight segments gives the discontinuity across the branch cut, and the divergence at the endpoints is canceled by the circular arcs.\label{keyhole}}
\centering
\end{figure}

\indent As reviewed in Section \ref{sec:BScorrections}, the Bohr-Sommerfeld quantization condition to first order in $1/\nu$ is 
\begin{align}
\nu \int_{z_-}^{z_+}dz\, (u^2- V(z))^{1/2}-\frac{1}{96i\nu}\oint dz\, \frac{V''(z)}{(V(z)-u^2)^{3/2}}=\pi(n+1/2),\hspace{10 mm}n\ge 0.\label{bscorrection}
\end{align}
Let us analyze this condition for a general potential $V(z)$, assuming that the potential has a minimum at $z=0$ and is monotonically increasing on either side of the minimum. The turning points are at $z_{\pm}$, where $z_{+}>0>z_-$. The contour integral is taken counterclockwise around the turning points. We are interested in small fluctuations near the minimum, so we expand the potential to quartic order, 
\begin{align}
V(z)=V(0)+\frac{1}{2!}V''(0) z^2+\frac{1}{3!}V'''(0)z^3+\frac{1}{4!}V''''(0)z^4+\ldots 
\end{align}
\indent The first integral can be evaluated in perturbation theory around $z=0$ \cite{2011arXiv1112.4247B}, 
\begin{align}
\int_{z_-}^{z_+} dz\, (u^2-V(z))^{1/2}=\frac{\pi(u^2-V)}{\sqrt{2V''}}\left(1+\frac{(u^2-V)(5V'''^2-3V''V''''')}{48V''^{3}}\right).
\end{align}
Here and below, the potential and its derivatives are evaluated at $z=0$. \\
\indent Now let us turn to the second integral in (\ref{bscorrection}). Note that this integral is divergent if integrated along the real axis from $z_-$ to $z_+$. Therefore we use a keyhole contour as in Figure \ref{keyhole}.
In practice what this means is that we integrate on the real axis from $z_-+\epsilon$ to $z_+-\epsilon$ and then discard divergent terms (which are cancelled by the circular contours near the singularities). Therefore we need to compute the finite part of 
\begin{align}
\frac{1}{48\nu}\int_{z_-+\epsilon}^{z_+-\epsilon}dz\, \frac{V''(z)}{ (u^2-V(z))^{3/2}}.
\end{align}
To fourth order we have 
\begin{align}
u^2=V+\frac{1}{2}V''z_{\pm }^2+\frac{1}{6}V'''z_{\pm}^3+\frac{1}{24}V''''z_{\pm}^4
\end{align} 
Expanding the integrand then gives
\begin{align}
&\frac{1}{48 \nu}\int_{z_-+\epsilon}^{z_+-\epsilon}dz\,  \frac{V''+V'''z+\frac{1}{2}V''z^2}{(\frac{1}{2}V''(z^2-z_{\pm}^2)+\frac{1}{6}V'''(z^3-z_{\pm}^3)+\frac{1}{24}V''''(z^4-z_{\pm}^4))^{3/2}}\notag\\
&\approx \frac{1}{12\sqrt{2}\nu V''^{3/2}}\int_{z_-+\epsilon}^{z_+-\epsilon}\frac{dz}{(z_{\pm}^2-z^2)^{3/2}} \left(V''+\frac{V'''(z^2+z_{\pm}z-z_{\pm}^2)}{2(z_{\pm}+z)}\right.\notag\\
&\hspace{30 mm}\left.+\frac{V'''^2(z_{\pm}^2+z_{\pm}z+z^2)(5z_{\pm}^2-7z_{\pm}z-7z^2)}{24V''(z_{\pm}+z)^2}+\frac{(3z^2-z_{\pm}^2)V''''}{8V''}\right).
\end{align}
\indent We split the integral into a piece from $z_-+\epsilon$ to zero and a piece from zero to $z_+-\epsilon$. In the first piece we replace $z_{\pm}$ by $z_-$ in the above expression, and in the second piece we replace $z_{\pm}$ by $z_+$. The integral evaluates to 
\begin{align}
\text{divergent}+\frac{1}{2\sqrt{2}\nu V''^{3/2}}\left(\frac{(21\pi+16)V'''^2}{432V''}-\left(\frac{1}{9z_-}+\frac{1}{9z_+}\right)V'''-\frac{\pi V''''}{16}\right).\label{integeval}
\end{align}
The divergent terms are canceled by the circular contours, and the remaining piece is finite. The final step is to solve for $z_{\pm}$ perturbatively in $V'''$. We get 
\begin{align}
z_\pm=\pm \sqrt{\frac{2(u^2-V)}{V''}}+\frac{(V-u^2)V'''}{3V''^2}+\ldots 
\end{align}
Terms proportional to $V''''$ and $V'''^2$ in $z_{\pm}$ do not contribute at this order and we can neglect them.
Expanding (\ref{integeval}) we get 
\begin{align}
\frac{\pi(7V'''^2-9V''V'''')}{288\sqrt{2}V''^{5/2}}.
\end{align}
\indent Collecting everything together, (\ref{bscorrection}) becomes
\begin{align}
\frac{1}{\sqrt{2V''}}\left(\nu(u^2-V)+\frac{\nu (u^2-V)^2(5V'''^2-3V''V''''')}{48V''^{3}}+\frac{7V'''^2-9V''V''''}{288\nu V''^{2}}\right)=n+1/2.\label{finalquant}
\end{align}
\section{Double-twist anomalous dimensions in general $d$}\label{gammagend}
In this appendix we collect the formulas in general dimensions for the anomalous dimensions to first order in $1/\nu$ and second order in $\mu$. We find
\begin{align}
    \gamma_1&=-\frac{1}{768k^{d/2}}\left[384k\nu+48 (2n+1)( d(d+2)k+(d-2)(d-4))\right.\\
    &\hspace{5 mm}\left.+\frac{d(d-2)}{k\nu}\left(6n(n+1)((d+2)(d+4)k^2+2(d+2)(d-4) k+(d-4)(d-6))\right.\right.\notag\\
    &\hspace{25 mm}\left.\left.+(d+2)(3d+4)k^2+6(d+2)(d-4) k+(d-4)(3d-10)\right)\right]+O(1/\nu^2) \ .\notag
\end{align}
This formula matches the result of \cite{Li:2020dqm} in 4, 6, 8, and 10 dimensions, see (4.16)-(4.18) in that paper. At the second order we find
\be
    \gamma_2&=-\frac{1}{18432k^d}\left[576(d^2k+d^2-4d+4)k\nu\right.\\
   &\left. \hspace{5 mm}+36((5d^4+4d^3-12d^2)k^2+(10d^4-40d^3+40d^2)k+5d^4-44d^3+132d^2-160d+64)(1+2n)\right.\notag\\
   &\left.+\frac{d(d-2)}{k\nu}\left(6n(n+1)((11d^4+40d^3-8d^2-88d)k^3+(33d^4-48d^3-108d^2+192d-96)k^2\right.\right.\notag\\
   &\left.\left.\hspace{20 mm}+(33d^4-216d^3+396d^2-240d)k+11d^4-128d^3+496d^2-712d+288)\notag\right.\right.\\
   &\hspace{20 mm}\left.\left.+(29d^4+64d^3-56d^2-136d)k^3+(87d^4-168d^3-180d^2+384d-96)k^2\right.\right.\notag\\
   &\hspace{20 mm}\left.\left.+(87d^4-528d^3+900d^2-432d)k+29d^4-296d^3+1024d^2-1336d+480\right)\right]+O(1/\nu^2).\notag
\ee
This result agrees with the one in \cite{Kulaxizi:2018dxo} in $d=2,4$ and $6$, see formula (6.39) in that paper.

\section{Bulk higher spin symmetry and the twist gap}
\label{app:bulkhigherspin}

The twist $\tau = \Delta - J$ spectrum of unitary CFTs in $d>2$ has a finite gap due to the unitarity bound $\tau \geq d-2$. In AdS this corresponds to the fact that interactions between a pair of objects go to zero at large separation (this is known as AdS clustering). One way to generate a separation between a pair of objects is by spinning them around each other. In this way large spin leads to large separation between a pair of objects, and thanks to the twist gap, the interaction between them goes to zero \cite{Alday:2007mf}. The same phenomenon can be understood by considering the light-cone limit of the crossing equation \cite{Fitzpatrick:2012yx,Komargodski:2012ek}, see also \cite{Alday:2015eya,Alday:2015ewa,Alday:2016njk,Simmons-Duffin:2016wlq,Caron-Huot:2017vep}. The punchline of this analysis is that the spectrum of interacting CFTs reproduces the one of the generalized free field (GFF), or, equivalently, the free field in AdS, at large spin and fixed twist.

It might be instructive to think about this phenomenon in terms of emerging symmetry. Let us consider the free scalar field in AdS. It admits an infinite set of conserved higher spin currents in the bulk, which schematically take the form, see e.g. \cite{Bekaert:2014cea},
\be
\cJ_J(x,z) =\sum_{k=0}^J c_k (z \cdot \nabla)^k \phi (z \cdot \nabla)^{J-k} \phi + ... ,
\ee
where $z^\mu$ is a polarization vector and $...$ stand for the trace terms. 

The bulk field $\phi$ is dual to the boundary operator $\cO$ of dimension $\Delta$. On the boundary the manifestation of \emph{bulk higher spin symmetry} is the statement that correlation functions of the GFF are invariant under the transformation
\be
[Q_J, \cO] = \partial^{J-1} \cO . 
\ee
Note that bulk higher spin symmetry is different from the CFT higher spin symmetry considered in \cite{Maldacena:2011jn,Maldacena:2012sf}. In the latter case the 
boundary theory contains a set of local conserved higher spin symmetry currents, which is not the case here. Indeed, in the case of the GFF the spectrum contains the operator $\cO$ as well as a set of multi-twist operators (schematically): $\cO \Box^n \partial^J \cO$, $\cO \Box^{n_1} \partial^{J_1} \cO \Box^{n_2} \partial^J_2 \cO$, ... . The bulk higher spin currents are then some objects made out of double-twist operators $\cO \Box^n \partial^J \cO$ with dimension
\be
\label{eq:doubletwistdim}
\Delta_{n,J} = 2 \Delta + 2n + J ,
\ee
and spin $J$. The presence of the equidistant operators \eqref{eq:doubletwistdim} then can be understood as a consequence of the bulk higher spin symmetry. The statement of AdS clustering, or the light-cone bootstrap, can be stated as an emergence of bulk higher spin symmetry at large spin.\footnote{In fact flat space clustering can be understood in the same way: in this case we can separately change the relative distance between two groups of widely separated particles/operators without changing the amplitude/correlator, see \cite{Coleman:1967ad} for a related discussion.} 

The presence of the bulk higher spin symmetry makes the theory integrable and leads to revivals and lack of thermalization. 
If such states persist in an interacting theory and are present amidst thermalizing states, they will serve as an example of many-body scars (a small fraction of non-thermalizing states in the otherwise thermalizing continuum) in higher-dimensional CFTs. As should be clear from our analysis the double-twist operators observed in the light-cone bootstrap analysis of the heavy-heavy-light-light four-point function are precisely such states, perturbatively in spin. For example, we showed in Section \ref{perturbativescars} that the one-point function in such states does not obey ETH. In this context bulk higher spin symmetry is called the spectrum-generating algebra \cite{serbyn,PhysRevB.102.085140}.
 
 On the other hand, as we have seen, this conclusion would be too fast due to the presence of important nonperturbative in spin effects (due to the presence of the black hole horizon) which are not captured by the naive light-cone bootstrap analysis. These effects remove the scar states from the spectrum by turning them into long-lived resonances comprised of the superposition of black hole microstates. A similar phenomenon was considered in the context of many-body scars in \cite{lin2020slow}.

\bibliographystyle{JHEP}
\bibliography{arXivfinalbib2}
\end{document}